\documentclass[12pt]{article}
\pdfoutput=1
\usepackage[a4paper]{geometry}
\usepackage{amsmath,amssymb,amsfonts,latexsym}
\usepackage{graphicx}  
\usepackage[colorlinks=true, pdfstartview=FitV, linkcolor=blue, citecolor=blue, urlcolor=blue]{hyperref}

\newcommand{\dS}{\mathbb S} 
\newcommand{\ppsi}{{\Psi}}  
\newcommand{\iz}{z^{-1}} 
\newcommand{\bra}{\langle} 
\newcommand{\ket}{\rangle}
\newcommand{\ep}{\qquad {\vrule height 10pt width 8pt depth 0pt}}
\newcommand {\bR}{{\mathbb R}}
\newcommand {\bN}{{\mathbb N}}
\newcommand {\bM}{{\mathbb M}}
\newcommand {\bZ}{{\mathbb Z}}
\newcommand {\bI}{{\mathbb I}}
\newcommand {\bJ}{{\mathbb J}}
\newcommand {\bC}{{\mathbb C}}

\newcommand {\bE}{{\mathbb E}}

\newcommand {\bT}{{\mathbb T}}
\newcommand {\bP}{{\mathbb P}}

\newcommand {\bsigma}{{\gimel}}

\newcommand {\cF}{{\cal F}}

\newtheorem{theorem}{Theorem} [section]
\newtheorem{lemma}[theorem]{Lemma}
\newtheorem{propo}[theorem]{Proposition}
\newtheorem{defi}[theorem]{Definition}
\newtheorem{corollary}[theorem]{Corollary}

\newtheorem {remark}[theorem]{Remark}

\begin{document}
\title{Localization Properties of the Chalker-Coddington Model\\
{\normalsize \it We dedicate this work to the memory of our friend and colleague Pierre Duclos} \bigskip }
\author{Joachim Asch
\thanks{
CPT-CNRS UMR 6207,
Universit\'e du Sud, ToulonÐVar, BP 20132,
F--83957 La Garde Cedex, France, e-mail:
asch@cpt.univ-mrs.fr},
Olivier Bourget
\thanks{
Departamento de Matem\'aticas
Pontificia Universidad Cat\'olica de Chile, Av. Vicu\~{n}a Mackenna 4860,
C.P. 690 44 11, Macul
Santiago, Chile},
Alain Joye
\thanks{
Institut Fourier, Universit\'e de Grenoble 1, BP 74,38402 Saint-Martin d'H\`eres,
France}
}
\date{30.07.2010}
\maketitle

\abstract{The Chalker Coddington quantum network percolation model is numerically pertinent to the understanding of the delocalization transition of the quantum Hall effect. We study the model restricted to a cylinder of perimeter $2M$.  We prove firstly that the Lyapunov exponents are simple and in particular that the localization length is finite; secondly that this implies spectral localization. Thirdly we prove a Thouless formula and compute the mean Lyapunov exponent which is independent of $M$.

\section{Introduction}

We start with a mathematical then a physical description of the model.
Fix the parameters
\[r, t\in\lbrack0,1\rbrack, \hbox{ \rm such that}, r^{2}+t^{2}=1,\]
denote by $\bT$ the complex numbers of modulus $1$ and for $q=(q_{1},q_{2},q_{3})\in\bT^{3}$, by $S(q)$ the general unitary $U(2)$ matrix depending on these three phases
\[S(q):=
\left(
\begin{array}{cc}
 q_{1}q_{2}& 0   \\
0&q_{1}\overline{q_{2}}     
\end{array}
\right)
\left(
\begin{array}{cc}
 t& -r   \\
r&t     
\end{array}
\right)
\left(
\begin{array}{cc}
 q_{3}& 0   \\
0&\overline{q_{3}}     
\end{array}
\right).
\]
Let $\left(\widehat{\Omega}, \widehat{\cF}, \widehat{\bP}\right)$ be the probability space: $\widehat{\Omega}:=\left(\bT^{6}\right)^{(2\bZ)^{2}}$, $\widehat{\bP}:=\otimes_{(2\bZ)^{2}} d^{6}l$ where $dl$ is the normalized Lebesgue measure on $\bT$, and $\widehat{\cF}$ the $\sigma-$algebra generated by the cylinder sets. With 
\[p\in\widehat{\Omega},\quad p(2j,2k)=:(\underbrace{p_{1},p_{2},p_{3}}_{p_{e}(2j,2k)}, \underbrace{p_{4}, p_{5}, p_{6}}_{p_{o}(2j+1,2k+1)})\]
and  the basis vectors $e_{\mu}(\rho):=\delta_{\mu, \rho}\quad (\mu, \rho\in\bZ^{2})$, the family of unitary operators
\[\widehat{U}(p): l^{2}(\bZ^{2})\to l^{2}(\bZ^{2})\]
is defined by its matrix elements $\widehat{U}_{\mu; \nu}=\bra e_{\mu}, \widehat{U}e_{\nu}\ket$: 

\[\widehat{U}_{\mu; \nu}:=0 \hbox{ \rm except for the blocks }\]
\begin{equation}\label{def:block}
\left(
\begin{array}{cc}
  \widehat{U}(p)_{(2j+1, 2k); (2j, 2k)} & \widehat{U}(p)_{(2j+1, 2k); (2j+1, 2k+1)}  \\
  \widehat{U}(p)_{(2j, 2k+1); (2j, 2k)}&    \widehat{U}(p)_{(2j, 2k+1); (2j+1, 2k+1)}   
\end{array}
\right):= S(p_{e}(2j,2k))
\end{equation}
\[
\left(
\begin{array}{cc}
  \widehat{U}_{(2j+2, 2k+2); (2j+2, 2k+1)} & \widehat{U}_{(2j+2, 2k+2); (2j+1, 2k+2)}  \\
  \widehat{U}_{(2j+1, 2k+1); (2j+2, 2k+1)}&    \widehat{U}_{(2j+1, 2k+1); (2j+1, 2k+2)}   
\end{array}
\right):= S(p_{o}(2j+1,2k+1)).
\]

Note that $\widehat{U}$ is an ergodic family of random unitary operators; indeed, $\widehat{U}^{\ast}\widehat{U}=\bI=\widehat{U}\widehat{U}^{\ast}$ because of the unitarity of the blocks; further
denote by $\widehat\Theta$ the action of $\bZ^{2}$ on functions $f$ on $\bZ^{2}$:
\[(\widehat\Theta_{(l,m)}f)(\mu):=f(\mu+(2l,2m))\qquad (\mu\in\bZ^{2}, (l,m)\in\bZ^{2}),\]
and, by abuse of notation, the corresponding shift on $
\widehat{\Omega}$. Then $\widehat\Theta$ is measure preserving and ergodic on $\widehat{\Omega}$ and \[\widehat{U}(\widehat\Theta p)=\widehat\Theta\widehat{U}(p)\widehat\Theta^{-1}.\]

This model was introduced in the physics literature by Chalker and Coddingtion, \cite{cc}, see \cite{kok} for a review, in order to study essential features of the quantum Hall transition in a quantitative way. $\widehat{U}$ describes the dynamics of a 2D electron in a strong perpendicular magnetic field and a smooth bounded random electric potential which is supposed to have some array of hyperbolic fixed points forming the nodes of a graph. 

In this picture the electron moves on the directed edges of the graph whose nodes are ``even'': $\{(1/2, 1/2)+(2j,2k), j,k\in\bZ\}$ or ``odd'': $\{(1/2, 1/2)+(2j+1,2k+1), j,k\in\bZ\}$ with edges connecting the even (odd) nodes to there nearest odd (even) neighbors. $\widehat{U}$ describes the evolution at time one of the electron. The edges are labeled by their midpoints. They are directed in such a way that $\widehat{U}$ models the tunneling near the hyperbolic fixed points of the potential, see figure \ref{fig:network}. The tunneling is described by the scattering matrices $S$ associated with the even, respectively odd, nodes. The i.i.d. random phases associated with each node take into account the deviation of the random electric potential from periodicity.

\begin{figure}[hbt]
\centerline {
\includegraphics[width=6cm]{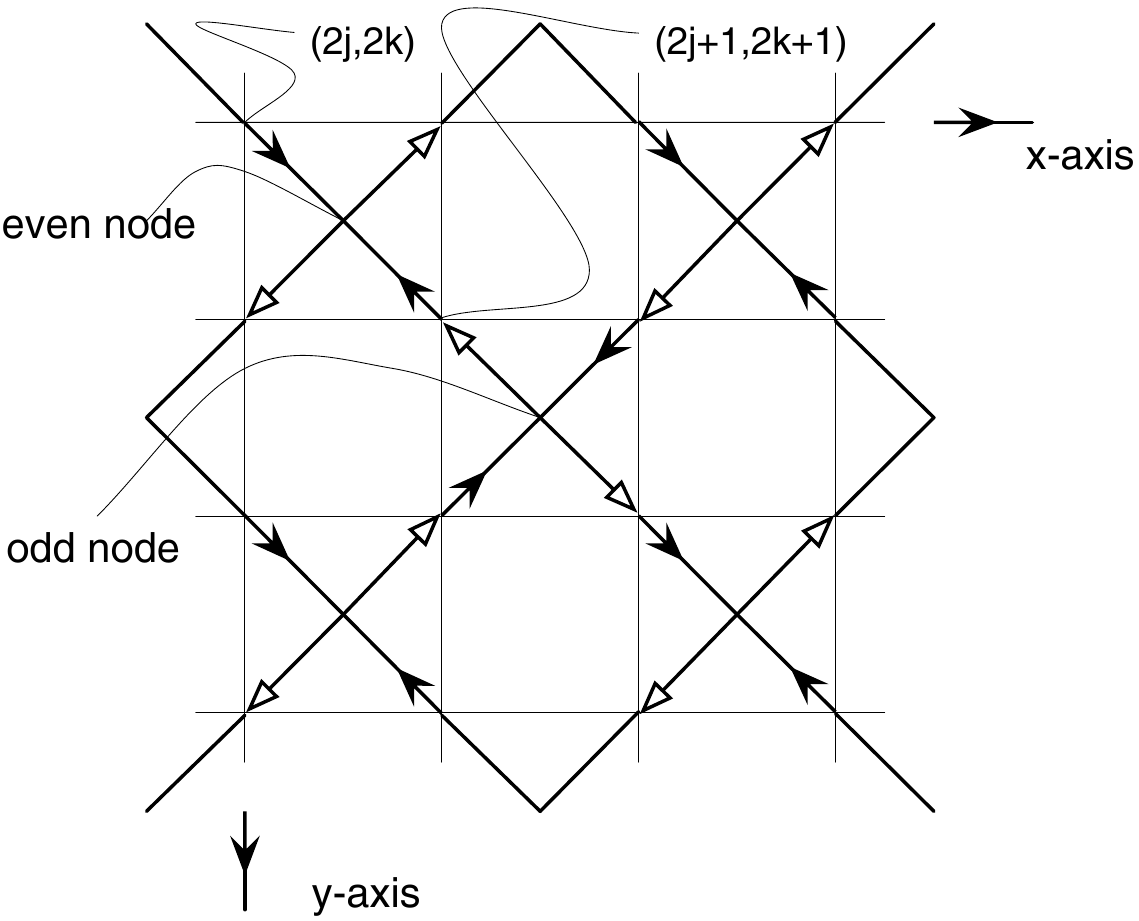}
}
\caption{The network model with its incoming (solid arrows) and outgoing links}
\label{fig:network}
\end{figure}

Following the literature on tunneling near a hamiltonian saddle point, \cite{fh}, \cite{cdvp1}, the parameter $t$ is $\frac{1}{\sqrt{1+e^{\varepsilon}}}$ where $\varepsilon$ is the distance of the electrons energy  to the nearest Landau Level. An application of  a finite size scaling method to their numerical observations led   Chalker and Coddington \cite{cc}, see also \cite{kok}, to conjecture that the localization length diverges as $t/r\to1$ as 
\[\left(\frac{1}{\ln\vert\frac{t}{r}\vert}\right)^{\alpha}\]
where the critical exponent  $\alpha$ exceeds substantially the exponent expected when a classical percolation model is applied to the problem, \cite{t}; the values advocated for $\alpha$ are $2.5\pm0.5$ for the quantum and $4/3$ for the classical case. 

Because of its importance for the understanding of the integer quantum Hall effect the one electron magnetic random model in two dimensions was and continues to be heavily studied in the mathematical literature. Mathematical results concerning the full Schr\"odinger Hamiltonian  can be traced from the following contributions and their references:  \cite{w2} for  percolation, \cite{gks}  for the existence of the localization--delocalization transition \cite{ass, besb,g} for the general theory of the quantum Hall effect. For results concerning a 2D electron in a magnetic field and periodic potential, which corresponds to the absence of phases here,  see \cite{tknn}, \cite{hs}. For recent work on Lyapunov exponents on hamiltonian strip models see \cite{rs}, \cite{bs}, \cite{bou}. 

Our results concern the restriction of the model to a strip of width $2M$ and periodic boundary conditions; they are presented as follows. In section \ref{sec:properties} we analyze the extreme cases, $r=0$ and $r=1$. Then, for the case where all phases are chosen to be 1, we  give a description  of the spectrum. Questions related to transfer matrix formalism are handled in sections \ref{sec:transfer}, \ref{sec:phases}, \ref{sec:exponents}  . In section \ref{sec:finiteness} we prove simplicity of the Lyapunov spectrum and finiteness of the localization length. In section \ref{sec:thouless} we prove a Thouless formula and show that the density of states is flat which implies our results on the mean Lyapunov exponent. 
In section \ref{sec:localization} we prove complete spectral localization.

\section{Some properties of the model}\label{sec:properties}
\subsection{Extreme cases}
Note that in case of complete ``reflection'' or ``transmission'' the system localizes completely: 
\begin{propo}\label{2d-pp} Let $rt=0$. Then, for any $p\in\widehat{\Omega}$, the spectrum of $\widehat{U}(p)$ is pure point.
\end{propo}
\noindent {\bf Proof:} Assume $r=0$, $p\in \widehat{\Omega}$ and define the family of subspaces $({\cal H}_{j,k})_{(j,k)\in {\mathbb Z}^2}$ as:
\[
{\cal H}_{j,k} = \mathrm{Ran} (e_{2j,2k},e_{2j+1,2k},e_{2j+1,2k-1},e_{2j,2k-1})\enspace.
\]
These subspaces are invariant under $\widehat{U}(p)$ and
\begin{equation}\label{eq:hblocks}
\oplus_{(j,k)\in {\mathbb Z}^2} {\cal H}_{j,k} = l^2({\mathbb Z}^2)\enspace,
\end{equation}
which means the operator $\widehat{U}(p)$ is pure point. The case $t=0$ is treated similarly. \ep

On the other hand one has complete propagation if all the phases are equal to one; define $\widehat{\Omega}\ni p=(\ldots, 1,1,1,\ldots)$ by $p(2j,2k):=(1,1,1,1,1,1)$ then we have:
{
\begin{propo}\label{propo:2d-ac} Let $rt\neq 0$. Then, the spectrum of $\widehat{U}(\ldots, 1, 1, 1, \ldots)$ is purely absolutely continuous. 
\end{propo}
\noindent {\bf Proof:} 
We make use of a  decomposition similar to (\ref{eq:hblocks}) and define the unitary 
 $V$ from $l^2({\mathbb Z}^2)$ to $l^2(\bZ^{2})\otimes \bC^{4}$ by
 $Ve_{2j,2k}:=e_{j,k}\otimes e_{1},Ve_{2j+1,2k+1}:=e_{j,k}\otimes e_{2},Ve_{2j,2k+1}:=e_{j,k}\otimes e_{3},Ve_{2j+1,2k}:=e_{j,k}\otimes e_{4}$. Let $P$ be the projection $P:=\bI\otimes\left(|e_{1}\ket\bra e_{1}|+|e_{2}\ket\bra e_{2}|\right)$. From the definition of $\widehat{U}$ in (\ref{def:block}) one reads that $V\widehat{U}^{2}V^{-1}$ commutes with $P$ and that $PV\widehat{U}^{2}V^{-1}P$ is equivalent to
 \[\left(
\begin{array}{cc}
rt (T_{0,1}-T_{1,0}) & r^2T_{1,0}+t^2T_{0,1}\\
t^2T_{0,-1}+r^2T_{-1,0} & rt(T_{-1,0}-T_{0,-1})
\end{array}\right)
 \]
 with the translations on $l^{2}(\bZ^{2})$ defined by 
 \[T_{n,m}\psi(j,k):=\psi(j+n,k+m)\qquad (n,m\in\bZ).\]
 The Fourier transform $\cF: l^{2}(\bZ^{2})\to L^{2}(\bT^{2})$ transforms the translations to multiplication operators: $\cF T_{n,m}\cF^{-1}=\exp{(-i(n x+m y))}$, thus the restriction to the range of $P$ of $\cF PV\widehat{U}^{2}V^{-1}P\cF^{-1}$ is equivalent to a matrix valued multiplication operator
 \begin{equation}\label{eq:matrixoperator}
 \left(
\begin{array}{cc}
rt (e^{-iy}-e^{-ix}) & r^2e^{-ix}+t^2e^{-iy}\\
t^2e^{iy}+r^2e^{ix}& rt(e^{ix}-e^{iy})
\end{array}\right).
 \end{equation}
 The trace of this matrix is not constant, its determinant is $-1$  hence the spectral bands are not flat, thus the spectrum of the restriction of $\widehat{U}^{2}$  is purely absolutely continuous. By an analogous argument this also holds for the restriction to $P^{\perp}$. \ep
}

Remark that a more general periodic distribution of phases leads to matrix valued translation operators with periodic coefficients thus to non-trivial Hofstadter like problems.

\section{Restriction to a cylinder, transfer matrices}\label{sec:transfer}
Let $M\in\bN$. Use the notation $\bZ_{2M}:=\bZ/(2M\bZ)$ for the discrete circle of perimeter  $2M$. Consider the restriction of the model to the cylinder $\bZ\times\bZ_{2M}$: 
\[U(p):l^{2}(\bZ\times\bZ_{2M})\to l^{2}(\bZ\times\bZ_{2M})\]
defined by its matrix elements with respect to  the canonical basis
\begin{equation}
U(p)_{\mu, \nu}:=\widehat{U}_{(\mu_{1},\mu_{2}\ mod\ 2M);(\nu_{1},\nu_{2}\ mod\ 2M)}.\label{def:U}
\end{equation}

Remark that $U(p)$ has the same spectral properties for some extreme cases as $\widehat{U}(p)$,  the model on the full lattice:

\begin{propo}\label{strip-pp} Let $rt=0$. Then, for any $p\in \Omega$, the spectrum of $U(p)$ is pure point.
\end{propo}
\noindent {\bf Proof:} Similar to the proof of Proposition \ref{2d-pp}. \ep

\begin{propo}\label{strip-ac} Let $rt\neq 0$. Then the spectrum of $U(\ldots, 1, 1, 1, \ldots)$ is purely absolutely continuous.
\end{propo}
\noindent {\bf Proof:} In the proof of proposition \ref{propo:2d-ac} note that $V$ now acts from $l^{2}(\bZ\times\bZ_{2M})$ to $\l^{2}(\bZ\times\bZ_{M})\otimes\bC^{4}$ and replace $\cF$ by the  Fourier transform from $l^{2}(\bZ\times\bZ_{M})$ to  $L^{2}(\bT\times\bZ_{M})$ defined by
\[\cF\psi(x,\kappa)=\sum_{j\in\bZ,k\in\bZ_{M}}\psi_{j,k}e^{i x j}e^{i \frac{2\pi}{M}\kappa k}\]
which diagonalizes the translations. Then, setting  $y=\frac{2\pi}{M}\kappa\quad(\kappa\in\bZ_{M})$, the matrix valued multiplication operator obtained in (\ref{eq:matrixoperator}) is understood as a family of matrix valued operators indexed over $\bZ_{M}$. The spectral bands are not flat by the same argument.
\ep

\medskip

From now on we restrict the discussion to the case
\[rt\neq0.\]

In the following $z$ denotes a complex number; also, unless otherwise stated, all indices in the second variable are to be understood $mod\ 2M$, e.g.: 
\[\psi_{2j, 2k+1}=\psi_{2j, 2k+1\ mod \ 2M}=\psi_{2j, 2k+1[2M]}.\]
A standard approach to the spectral problem of $U$ is the transfer matrix method. Though this is well known, we wish to recall the construction explicitly for the model at hand. 
\begin{propo}\label{propo:tMatrices}
For $z\neq0$, $q=\left(q_{1},q_{2},q_{3}\right)\in\bT^{3}$ define
\[
T_{eo}(z,q):=\left(
\begin{array}{cc}
  q_{1}q_{2}&0      \\
  0& q_{3}        
\end{array}
\right)
\frac{1}{t}\left(
\begin{array}{cc}
 \iz & -r      \\
  -r& z        
\end{array}
\right)
\left(
\begin{array}{cc}
  q_{3}&0      \\
  0& \overline{q_{1}}q_{2}        
\end{array}
\right),
\]
\[
T_{oe}(z,q):=\left(
\begin{array}{cc}
  \overline{q_{3}}&0      \\
  0& q_{1}q_{2}        
\end{array}
\right)
\frac{1}{r}\left(
\begin{array}{cc}
 z&-t      \\
  t& -\iz         
\end{array}
\right)
\left(
\begin{array}{cc}
  \overline{q_{1}}q_{2}&0      \\
  0& \overline{q_{3}}       
\end{array}
\right).
\]
Then 
\begin{enumerate}
\item For $\psi: \bZ\times\bZ_{2M}$ it holds:
\[\sum_{\nu\in\bZ\times\bZ_{2M}}U_{\mu \nu}\psi_{\nu}=z\psi_{\mu}\qquad \forall\mu\in\bZ\times\bZ_{2M}\]
\[\Longleftrightarrow\]
\[
\left(
\begin{array}{c}
  \psi_{2j+1,2k }    \\
  \psi_{2j+1,2k+1 }
 \end{array}
\right)
=T_{eo}\left(z,p_{e}(2j, 2k)\right)
\left(\begin{array}{c}
  \psi_{2j,2k }    \\
  \psi_{2j,2k+1 }
 \end{array}
\right)
\] 
and
\[
\left(
\begin{array}{c}
  \psi_{2j+2,2k+1 }    \\
  \psi_{2j+2,2k+2 }
 \end{array}
\right)=
T_{oe}\left(z,p_{o}(2j+1, 2k+1)\right)
\left(\begin{array}{c}
  \psi_{2j+1,2k+1 }    \\
  \psi_{2j+1,2k+2 }
 \end{array}
\right).
\] 
\item For $z\in\bT$, it holds that $T_{oe}, T_{eo}\in U(1,1)$,   the Lorentz group defined as a subset of the complex $2\times2$ matrices by
\[U(1,1):=\left\lbrace B\in\bM_{2,2}(\bC); B^{\ast}JB=J, \quad J:=\left(\begin{array}{cc} 1&0   \\0&-1 \end{array}\right)\right\rbrace
\]
\end{enumerate}

\end{propo}

{\bf Proof:} 

By definition of $U$ we have for the ``even'' nodes:
\[
\left(
\begin{array}{c}
  {(U\psi)}_{2j+1,2k}    \\
 (U \psi)_{2j,2k+1}
 \end{array}
\right)=
S\left(p_{e}(2j, 2k)\right)
\left(\begin{array}{c}
  \psi_{2j,2k}    \\
  \psi_{2j+1,2k+1}
 \end{array}
\right)=z\left(
\begin{array}{c}
  {\psi}_{2j+1,2k}    \\
  \psi_{2j,2k+1}
 \end{array}
\right),
\] 
and, for the ``odd'' nodes: 
\[
S\left(p_{o}(2j+1, 2k+1)\right)
\left(\begin{array}{c}
  \psi_{2j+2,2k+1}    \\
  \psi_{2j+1,2k+2}
 \end{array}
\right)=z\left(\begin{array}{c}
  {\psi}_{2j+2,2k +2}    \\
  \psi_{2j+1,2k+1}
 \end{array}\right).
\] 
For a matrix 
\[S=\left(
\begin{array}{cc}
 S_{11}&S_{12}     \\
  S_{21}&S_{22}        
\end{array}\right)\qquad\hbox{\rm with } S_{22}S_{21}\neq0
\]
it holds:

\[\left(\begin{array}{c} a   \\b\end{array}\right)=S\left(\begin{array}{c} x   \\y\end{array}\right)
\Longleftrightarrow
\left(\begin{array}{c} a   \\y\end{array}\right)=\widehat{S}\left(\begin{array}{c} x   \\b\end{array}\right)
\Longleftrightarrow
\left(\begin{array}{c} x   \\a\end{array}\right)=\breve{S}\left(\begin{array}{c} b   \\y\end{array}\right)
\]
with
\[
\widehat{S}=\frac{1}{S_{22}}\left(
\begin{array}{cc}
 \det{S}&S_{12}     \\
  -S_{21}&1        
\end{array}\right),\qquad\breve{S}=\frac{1}{S_{21}}\left(
\begin{array}{cc}
1&-S_{22}     \\
S_{11}&- \det{S}        
\end{array}\right).
\]

Now
\[
\left(
\begin{array}{cc}
t&-r     \\
r&t        
\end{array}\right)^{\widehat{}}=
\frac{1}{t}\left(
\begin{array}{cc}
1&-r    \\
  -r&1        
\end{array}\right);
\quad \left(
\begin{array}{cc}
t&-r     \\
r&t        
\end{array}\right)^{\breve{}}=
\frac{1}{r}\left(
\begin{array}{cc}
1& -t  \\
  t&-1        
\end{array}\right)
\]
so
\[z\left(\begin{array}{c}
a\\
b        
\end{array}\right)=
q_{1}\left(\begin{array}{cc}
q_{2}& 0  \\
  0&\overline{q_{2}}       
\end{array}\right)
\left(\begin{array}{cc}
t& -r  \\
  r&t        
\end{array}\right)
\left(\begin{array}{cc}
q_{3}& 0  \\
  0&\overline{q_{3}}        
\end{array}\right)
\left(\begin{array}{c}
x\\
 y        
\end{array}\right)\]
\[\Longleftrightarrow \left(\begin{array}{c}
a\\
y        
\end{array}\right)=
\left(\begin{array}{cc}
q_{1}q_{2}& 0  \\
  0&q_{3}       
\end{array}\right)
\frac{1}{t}\left(\begin{array}{cc}
\iz & -r  \\
 -r&z        
\end{array}\right)
\left(\begin{array}{cc}
q_{3}& 0  \\
  0&\overline{q_{1}} q_{2}       
\end{array}\right)
\left(\begin{array}{c}
x\\
 b        
\end{array}\right)\]
\[\Longleftrightarrow \left(\begin{array}{c}
x\\
a        
\end{array}\right)=
\left(\begin{array}{cc}
\overline{q_{3}}& 0  \\
  0&q_{1}q_{2}       
\end{array}\right)
\frac{1}{r}\left(\begin{array}{cc}
z& -t  \\
t&-\iz         
\end{array}\right)
\left(\begin{array}{cc}
\overline{q_{1}} q_{2} & 0  \\
  0&   \overline{q_{3}}   
\end{array}\right)
\left(\begin{array}{c}b\\y       \end{array}\right)\]
from which the first  claim follows.

Denote by $\bI$ the identity matrix in $\bC^{2}$. $S$ is a unitary matrix if and only if the pullback of the quadratic form in $\bC^{4}$ associated with $Q=\left(
\begin{array}{cc}\bI   &   \\  &  -\bI     \end{array}\right)$   (blanks stand for $0$ entries) to the graph of $S$: $\left\{(u,Su)\in\bC^{4}, u\in\bC^{2}\right\}$ is zero. The mapping from $(x,y,a,b)$ to $(x,b,a,y)$ transforms $Q$ to $\left(
\begin{array}{cc}J   &   \\  &  -J    \end{array}\right)$. The pullback of the corresponding form to the graph of $T_{eo}$ being zero, it follows that  $T_{eo}$ and, by the analogous argument,   $T_{oe}$,  belong to the Lorentz group.
\ep

For later use we fix the following notation
\begin{defi}
Denote by $\bJ$ the $2M\times 2M$ block diagonal matrix consisting of $M$ non-zero diagonal blocks equal to $J$ and by
\[U_{M}(1,1):=\{B\in \bM_{2M,2M}(\bC); B^{\ast}\bJ B=\bJ \}.\]
 the unitary  group of the hermitian form defined by $\bJ$.
\end{defi}
Note that $U_{M}(1,1)$ is isomorphic to the classical unitary group $U(M,M)$ of the hermitian form $ \vert z_{1}\vert^{2} +\ldots+\vert z_{M}\vert^{2} -\vert z_{M+1}\vert^{2}\ldots-\vert z_{2M}\vert^{2}$.

\section{Relevant phases}\label{sec:phases}
Because of the uniform distribution  it is possible to reduce the number of relevant phases in the model to two phases per node. Before proceeding we do this reduction.
We shall repeatedly make use of 

\begin{lemma}\label{lemma:super}
Let $\varphi_1, \cdots, \varphi_n$ be independent and uniformly distributed random variables on  $\bR/\bZ$ and let $A\in \bM_{m,n}(\bZ)$. Then, $\theta_1, \cdots, \theta_m$ defined by $\vec \theta=A\vec \varphi$ are independent and uniformly distributed if and only if Rank $A$ is maximal.
\end{lemma}
{\bf Proof.} 
For $\vec k\in \bZ^m$ it holds
\[
\bE (e^{i\langle \vec k, \vec \theta \rangle})=
\bE (e^{i\langle \vec k, A \vec \varphi\rangle})=\bE (e^{i\langle A^t \vec k, \vec \varphi\rangle})=\delta_{A^t \vec k, 0}.
\]
Thus the $\vec\theta$ are independent and uniformly distributed if and only if $\bE (e^{i\langle \vec k, \vec \theta \rangle})=\delta_{\vec k, 0}$ if and only if  Ker$A^t=\{0\}$,  equivalently, if and only if Rank $A$ is maximal. \ep

\begin{propo}
There exists $g:\widehat{\Omega}\to\bT^{\bZ^{2}}$ such that for $p\in\widehat{\Omega}$ the evolution $\widehat{U}(p)$ defined by (\ref{def:block}) is unitarily equivalent to 
\[D(g(p)) \dS \hbox{ on } l^{2}(\bZ^{2})\]
where $D(q)$ is diagonal, $D(q)_{(j,k);(j,k)}=q_{j,k}$, and  $\dS=\widehat{U}(\ldots,1,1,1,\ldots)$. Moreover, the image measure  of  $\otimes_{(2\bZ)^{2}}d^{6}l$ by $g$ is $\otimes_{\bZ^{2}}dl$.
\end{propo}

{\bf Proof:}  By (\ref{def:block}), $\widehat{U}(p)$ is of the form $\widehat{U}(p)=D^{(1)}(p)\dS D^{(2)}(p)$ where $D^{(j)}(p)$ are diagonal, and defined by their diagonal elements:
\[
\begin{array}{ll}
D^{(1)}(p)_{2j+1, 2k}=p_1p_2(2j,2k), & D^{(1)}(p)_{2j, 2k+1}=p_1\bar{p}_2(2j,2k),\\
D^{(1)}(p)_{2j+2, 2k+2}=p_4p_5(2j+1,2k+1), & D^{(1)}(p)_{2j+1, 2k+1}=p_4\bar{p}_5(2j+1,2k+1),\\
D^{(2)}(p)_{2j, 2k}=p_3(2j,2k), & D^{(2)}(p)_{2j+1, 2k+1}=\bar{p}_3(2j,2k),\\
D^{(2)}(p)_{2j+2, 2k+1}=p_6(2j+1,2k+1), & D^{(2)}(p)_{2j+1, 2k+2}=\bar{p}_6(2j+1,2k+1).
\end{array}
\]
Hence, $\widehat{U}(p)$ is unitarily equivalent to  $D^{(2)}(p)D^{(1)}(p)\dS$ which has the asserted shape. Define $q=g(p)$ by
\begin{eqnarray*}
&&q(2j+1,2k):= \overline{p}_{6}(2j+1,2k-1)p_{1}p_{2}(2j,2k),\\
&&q(2j,2k+1):= p_{6}(2j-1,2k+1)p_{1}\bar{p}_{2}(2j,2k),\\
&&q(2j+2,2k+2):= p_{3}(2j+2,2k+2)p_{4}p_{5}(2j+1,2k+1),\\
&&q(2j+1,2k+1):= \overline{p}_{3}(2j,2k)p_{4}\bar{p}_{5}(2j+1,2k+1).
\end{eqnarray*}
Now, an application of Lemma \ref{lemma:super} shows the $q's$ are i.i.d. and uniformly distributed.
\hfill\ep

\begin{remark}
Note that the unitary transformation just constructed is diagonal and thus does not affect the localization properties of the model.
\end{remark}
In the following, we abuse notations and call  for $q\in\bT^{\bZ^{2}}$ the matrix operator  $D(q)\dS$ again $\widehat{U}(q)$; same abuse for the restriction to the cylinder.

\section{Characteristic exponents}\label{sec:exponents}
We now define and analyze the transfer matrices and in particular the localization length. Consider

 \begin{equation}
U(p)=D(p)\dS \quad\hbox{on } l^{2}(\bZ\times\bZ_{2M})\label{def:U2}
\end{equation}
with identically distributed uniformly distributed phases in
$\bT^{\bZ\times\bZ_{2M}},\otimes_{\bZ\times\bZ_{2M}}dl$ and the cylinder set algebra.

 We use the unitary equivalence
\begin{eqnarray}
l^{2}(\bZ\times\bZ_{2M})\cong l^{2}(\bZ\times\{0,\ldots, 2M-1\})&\to& l^{2}(\bZ;{l^{2}(\bZ_{2M})})\cong l^{2}(\bZ,{\bC^{2M}})\nonumber\\
\psi &\mapsto&\ppsi\label{def:ppsi}\\
\ppsi_{j}&:=&(\psi_{j,0}, \ldots,\psi_{j, 2M-1}).\nonumber
\end{eqnarray}

Note that with the reduced phases the building blocks of the transfer matrices read with phases {\it p,q}

\[
\left(
\begin{array}{cc}
  {\it p}&0      \\
  0& 1        
\end{array}
\right)
\frac{1}{t}\left(
\begin{array}{cc}
 \iz & -r      \\
  -r& z        
\end{array}
\right)
\left(
\begin{array}{cc}
  1&0      \\
  0& {\it q}       
\end{array}
\right),
\]
\[
\left(
\begin{array}{cc}
  1&0      \\
  0& {\it p}       
\end{array}
\right)
\frac{1}{r}\left(
\begin{array}{cc}
 z&-t      \\
  t& -\iz         
\end{array}
\right)
\left(
\begin{array}{cc}
  {\it q}&0      \\
  0& 1       
\end{array}
\right).
\]

As we shall explain below, the previous analysis leads us to deal with the following random dynamical system:

Consider  the probability space defined by $\Omega={(\bT^{4M})}^{2\bZ}$, $\bP=\otimes_{\bZ}d^{4M}l$, and $\cF$ the cylinder set algebra. 
The shift 
\[\Theta:\Omega\to\Omega, \qquad \Theta p(2m):=p(2(m+1))\qquad (m\in\bZ)\] is measure preserving and ergodic. For $p\in\Omega$ define the following elements of $\left(\bT^{2M}\right)^{2\bZ}$
\begin{eqnarray*}
&&p_{r}=(1,p_{1},1,p_{3},\ldots,1,p_{2M-1})\\
&&p_{l}=(p_{0},1,p_{2},1,\ldots,p_{2M-2},1)\\
&&p_{m}=(p_{2M},p_{2M+1},\ldots,p_{4M-1}).\end{eqnarray*}

Denote for $q\in\bT^{2M}$ the unitary diagonal matrix
\[D(q):=
\left(
\begin{array}{ccc}
  q_{1} \\
 & \ddots  \\
 &  & q_{2M}
\end{array}
\right), 
\]
(where  $0$ valued matrix entries are represented by blanks) and for $z\neq0$
 the $2M\times 2M$ matrices
 
\[M_{1}(z):=\frac{1}{t}\left(
\begin{array}{ccccc}
 \iz  &-r   \\
  -r&z \\
  &&\ddots  \\
  &&&\iz &-r\\
  &&&-r&z
  \end{array}
\right),
\]
\[M_{2}(z):=\frac{1}{r}\left(
\begin{array}{ccccccc}
 -\iz  &   & &&&&t   \\
  &  z & -t&&&&\\
  &t&-\iz  \\
  &&&\ddots  \\
  &&&&z&-t\\
  &&&&t&-\iz \\
 - t&&&&&&z  
\end{array}
\right).
\]
Define for a fixed $z\neq0$
\begin{eqnarray}
A_{z}:\Omega&\to &U_{M}(1,1)\nonumber\\
A_{z}(p)&:=&D\left(p_{l}\right)M_{2}(z)
D\left(p_{m}\right)M_{1}(z)D\left(p_{r}\right)\label{def:generator}.
\end{eqnarray}

Then $A$ generates the cocycle  $\Phi$ over the ergodic dynamical system
\[\left(\Omega,\cF,\bP, (\Theta^{n})_{n\in\bZ}\right)\]
defined by
$\Phi_{z}:\bZ\times\Omega\to U_{M}(1,1)$ 
\[\Phi_{z}(n,p):=
\left\lbrace
\begin{array}{ll}
  A(\Theta^{n-1}p)\ldots A(p)&n>0      \\
  \bI& n=0     \\
  A^{-1}(\Theta^{n}p)\ldots A^{-1}(\Theta^{-1}p)&n<0     
\end{array}
\right. .
\]
Oseledets theorem holds for $\Phi$, see \cite{a}, Theorem 3.4.11 and Remark 3.4.10 (ii):

\begin{defi}\label{def:exponents} Let $z\neq0$. There exists an invariant subset of full measure of $p\in\Omega$ such that the limits
\[\lim_{n\to\infty}\left(\Phi_{z}^{\ast}(n,p)\Phi_{z}(n,p)\right)^{1/2n}=\lim_{n\to-\infty}\left(\Phi_{z}^{\ast}(n,p)\Phi_{z}(n,p)\right)^{1/2\vert n\vert}=:\Psi_{z}(p)\]
exist. Denote by $\gamma_{k}(p,z)\quad k\in\{1,\ldots,2M\}$ the eigenvalues of $\Psi_{z}(p)$ arranged in decreasing order. Due to ergodicity there exists $\gamma_{k}(z)\ge0$ such that $\gamma_{k}(p,z)=\gamma_{k}(z)$ on an invariant subset of full measure. 
The characteristic exponents are defined by $\lambda_{k}(z):=\log{\gamma_{k}(z)}$. 
\end{defi}

Due to the Lorentz symmetry  of the transfer matrices for $z\in\bT$ we have
\begin{propo}
\begin{enumerate}
\item Let $B\in U_{M}(1,1)$. Then for the singular values $SV(B)$ it holds
\[\gamma\in SV(B)\Longleftrightarrow \frac{1}{\gamma}\in SV(B).\]
\item For $\lambda_{j}:=\log{\gamma_{j}},\quad\gamma_{j}\in SV(B)$ arranged in decreasing order it holds:
\[\lambda_{j+M}=-\lambda_{M-j+1}\qquad\forall j\in\{0,\ldots, M\}.\]
\end{enumerate}

\end{propo}

{\bf Proof:}  We have
$B^{\ast}\bJ B$=$\bJ$. In particular $\det{B}\neq0$, so $\gamma\neq0$ and \\ $\bJ^{-1}B^{\ast}=B^{-1}\bJ^{-1}$ as well as $B\bJ=\bJ{B^{\ast}}^{-1}$.  Now
\[ \det{\left(B^{\ast}B-z^{2} \right)}=0\Longleftrightarrow \det{\left(\bJ^{-1}B^{\ast}B\bJ-z^{2} \right)}=0\Longleftrightarrow\]
\[\det\left((B^{\ast}B)^{-1}-z^{2}\right)=0\Longleftrightarrow z^{4M}\det (B^{\ast}B)^{-1}\det\left(\frac{1}{z^{2}}-B^{\ast}B\right)=0.\]
From which the two claims follow.
\hfill\ep

Thus we restrict our discussions to the first $M$ non-negative Lyapunov exponents 
\[\lambda_{1}\ge\lambda_{2}\ge\ldots\ge\lambda_{M}\ge0\]
which we shall call for simplicity ``the'' Lyapunov exponents in the sequel.

We show that due to the translation invariance of the uniform distribution, the exponents are independent of $z$: 
\begin{lemma}\label{lemma:zinvariance}
For any $w\in \bT$,
\[A_{wz}(p)=A_z(w\odot p),\]
where $w\odot p$ is defined by $w\odot p_{2j}:=w^{-1}p_{2j}$, and $w\odot p_{2j+1}:=w p_{2j+1}$. 
\end{lemma}

{\bf Proof: } Write $D(p_m)$ in (\ref{def:generator}) as
$D(p_l')D(p_r')$, where 
\begin{eqnarray*}
&&p_{r}':=(1,p_{2M+1},1,p_{2M+3},\ldots,1,p_{4M-1})\\
&&p_{l}':=(p_{2M},1,p_{2M+2},1,\ldots,p_{4M-2},1). \end{eqnarray*} Thus $A_{z}(p)$ is the product of the block diagonal matrices
$D(p_l)M_2(z)D(p_l')$ and $D(p_r')M_1(z)D(p_r)$ whose blocks are
\begin{eqnarray*}
a_z^l(p,q)&:=&\left(\begin{array}{cc}
1&  0 \\
0&p         
\end{array}\right)\frac{1}{r}\left(\begin{array}{cc}
z& -t  \\
t&-\iz         
\end{array}\right)\left(\begin{array}{cc}
1& 0  \\
0&q         
\end{array}\right)=\frac{1}{r}\left(\begin{array}{cc}
z& -qt  \\
pt&-pq\iz         
\end{array}\right) \\
a_z^r(p,q)&:=&\left(\begin{array}{cc}
1&  0 \\
0&p         
\end{array}\right)\frac{1}{t}\left(\begin{array}{cc}
\iz& -r  \\
-r&z         
\end{array}\right)\left(\begin{array}{cc}
1& 0  \\
0&q         
\end{array}\right)=\frac{1}{r}\left(\begin{array}{cc}
\iz& -qr  \\
-pr&pqz         
\end{array}\right).
\end{eqnarray*}
For any $w\in \bT$, these matrices satisfy 
$$
a_{wz}^l(p,q)=w a_{z}^l(w^{-1}p,w^{-1}q), \ \ \ a_{wz}^r(p,q)=w^{-1} a_{z}^l(wp,wq),
$$
from which the result follows.\hfill \ep

Therefore, for any fixed $w\in \mathbb T$, the matrices $A_z(w\odot p)$ have the same distribution as $A_{wz}(p)$. As a consequence
\begin{corollary}\label{lemma:zindep}
All characteristic exponents $\lambda_k(z)=\lambda_k$ are independent of $z\in \mathbb T$.
\end{corollary}

{\bf Proof.}  $\lambda_{k}=\bE\left(\log(\gamma_{k}(z,p))\right)=\bE\left(\log(\gamma_{k}(1,\frac{1}{z}\odot p))\right)=\bE\left(\log(\gamma_{k}(1,p))\right)$.
\hfill\ep

\begin{defi}
The localization length $\xi_{M}\in\lbrack0,\infty\rbrack$ is defined as
\[\xi_{M}:=\frac{1}{\lambda_{M}}.\]
\end{defi}

\begin{remark}
In the physics literature,  see \cite{kok},  $\xi_{M}$ in assumed to be finite for all parameters; a change of the asymptotic behavior  as $M\to\infty$ is conjectured when  the parameters of the model approach the critical point $t=r$. This conjecture is supported by a numerical finite size scaling method and is supposed to reflect the divergence of the localization length of the full system at the critical point. Thus a first step to support these heuristics is to prove finiteness of $\xi_{M}$ and to establish precise information of its behavior as a function  of $M$.
\end{remark}

The announced equivalence to the propagation problem is the content of the following 

\begin{propo} \label{propo:eq}Let $U(p)$ be the ergodic family of unitary operators defined in (\ref{def:U2}) over the probability space $\Gamma:=\bT^{\bZ\times\bZ_{2M}},\otimes_{\bZ\times\bZ_{2M}}dl$ and the cylinder set algebra. Let $f: \Gamma\to\Omega$ be defined for $j\in\bZ$ by
\begin{eqnarray*}
f(p)(2j)&:=&\left(p_{2j,0},p_{2j+2,1},p_{2j,2},p_{2j+2,3}\ldots p_{2j,2M-2},p_{2j+2,2M-1}\right.\\
&&\left.p_{2j+1,0},p_{2j+1,1}\ldots p_{2j+1,2M-1}\right).
\end{eqnarray*}

The image measure by $f$ is the measure on $\Omega$ and it holds
 \[U(p)\psi=z\psi \Longleftrightarrow \ppsi_{2N}=\Phi_{z}(N,f(p))\ppsi_{0} \] 
for $\ppsi$ defined in (\ref{def:ppsi}).
\end{propo}

{\bf Proof.} The construction of $f$ follows from Proposition (\ref{propo:tMatrices}). The image measure follows from lemma (\ref{lemma:super}).
\ep

\section{Finiteness of the localization length}\label{sec:finiteness}
Using the methods exposed in \cite{bl}, see also \cite{gm}, we prove that all Lyapunov exponents are distinct and in particular that the localization length for the cylinder in finite.

\begin{theorem}\label{thm:finite}
For  $rt\neq0, z\in\bT$ it holds
\[\lambda_{1}>\lambda_{2}>\ldots>\lambda_{M}>0.\]
\end{theorem}

{\bf Proof.} We follow the strategy exposed in \cite{bl} and prove the theorem in several steps making use of lemmata to be proven below. Denote by
\[G:=\hbox{\rm the smallest subgroup of }U_{M}(1,1) \hspace{5pt}\hbox{\rm  generated by } \{A(p), p\in\Omega\}.\]
By lemma \ref{lem:fullgroup}
\[G=U_{M}(1,1).\]
In particular it is then known :
\[G\hspace{5pt}\hbox{\rm is connected}.\]

Furthermore, see also \cite{rs},  $G$ is isomorphic to the complex symplectic group. Indeed : denote by $\bsigma$ the $2M\times 2M$ block diagonal matrix consisting of $M$ non zero blocks $\sigma=\left(\begin{array}{cc}0 & -1  \\  1& 0  \end{array}\right)$; we write: $\bsigma=\oplus_{1}^{M}\sigma$ for short; denote by
\[Sp(M,\bC):=\{B\in\bM_{2M,2M}(\bC); B^{\ast}\bsigma B=\bsigma\}\]
the complex symplectic group. From
\[\frac{1}{\sqrt{2}}\left(\begin{array}{cc}1& -i  \\  1& i  \end{array}\right)^{\ast}J\frac{1}{\sqrt{2}}\left(\begin{array}{cc}1& -i  \\  1& i  \end{array}\right)=i\sigma\]
 it follows defining  $C:=\bigoplus_{1}^{M}\frac{1}{\sqrt{2}}\left(\begin{array}{cc}1& -i  \\  1& i  \end{array}\right)$ that
\[G=U_{M}(1,1)=C Sp(M,\bC) C^{\ast}.\]
In order to freely use  results in \cite{bl} we shall do our argument for real matrices. To this end we separate real and imaginary parts and  consider
\[\tau:\bM_{2M,2M}(\bC)\to\bM_{4M,4M}(\bR)\]
\[x=a+ib\mapsto\left(\begin{array}{cc}a& -b  \\  b& a \end{array}\right).\]
It holds: $\tau(x+y)=\tau(x)+\tau(y); \tau(xy)=\tau(x)\tau(y); \tau(x^{\ast})=\tau(x)^{t};ker\{\tau\}=\{0\}; \det{\tau(x)}=\vert\det{x}\vert^{2}$, thus

\[\tau\left(C^{\ast}GC\right)\subset Sp(2M,\bR)\]
with the real symplectic group

\[Sp(2M,\bR):=\{B\in\bM_{4M,4M}(\bR); B^{t}\bsigma B=\bsigma\}\]
for $\bsigma=\oplus_{1}^{2M}\sigma$.

As $\det{\tau(x)}=\vert\det{x}\vert^{2}$ implies that $\tau(x)$ shares its eigenvalues with $x$ with  the degeneracies  doubled. So the Lyapunov exponents $\gamma$ defined by the $\tau-C$ transformed products of transfer matrices are
\[\gamma_{1}=\gamma_{2}=\lambda_{1}\ge\ldots\ge\gamma_{2p-1}=\gamma_{2p}=\lambda_{p}\ge\ldots\ge\gamma_{2M-1}=\gamma_{2M}=\lambda_{M}.\]
As $\tau\left(C^{\ast}GC\right)$ is connected one can infer from  \cite{bl} Theorem 3.4 and Exercice 2.9 for $p\in\{1,\ldots,M\}$:

\[\left.\begin{array}{ll}\tau\left(C^{\ast}GC\right)& L_{2p}\hspace{5pt} \hbox{\rm  irreducible}  \\ and&\\ \tau\left(C^{\ast}GC\right)& \hspace{5pt} \hbox{\rm 2p contracting} \end{array}\right\}\Longrightarrow\gamma_{2p}=\lambda_{p}>\lambda_{p+1}=\gamma_{2p+1}\]
 in particular for $p=M$:  $\lambda_{M}>0$.
 Now by lemma \ref{lem:irreducible} and lemma \ref{lem:contracting} the group $\tau\left(C^{\ast}GC\right)$ is 2p irreducible and 2p contracting for all $p\in\{1,\ldots,M\}$ so all Lyapunov exponents are distinct and $\lambda_{M}>0$.
\ep

The following lemmata complete the proof of theorem \ref{thm:finite}, we use the notations introduced in the above proof.

\begin{lemma}\label{lem:fullgroup}
\[G=U_{M}(1,1).\]
\end{lemma}

{\bf Proof.} 
By definition $G\subset U_{M}(1,1)$ is a closed subgroup of $Gl(2M,\bC)$ thus $G$ is a Liegroup. By connectedness of $U_{M}(1,1)$ it is sufficient to show that the Lie algebras ${\mathfrak g}$ and ${\mathfrak u}_{M}(1,1)$ coincide. Now
\[{\mathfrak u}_{M}(1,1)=\{A\in \bM_{2M,2M}(\bC);   A_{jk}=-\overline{A}_{kj}(-1)^{k+j}\}\]
whose dimension as a real vector space equals $4M^{2}$.

Denote by  $D_j(t)=\mbox{diag}(1, 1, \dots, 1, e^{it}, 1, \dots, 1) $ the unitary matrix where the phase sits at the $j$'th slot, for $j=1, 2, \dots, 2M$ and use the $M_{j}$ as defined in section \ref{sec:exponents}. For $z\in\bT$ the matrices
\begin{equation}\label{eq:generators}
i |j\ket\bra j|, \ \ \  i M_2(z) |j\ket\bra j| M_2(z)^{-1},   \ \ \ i M_1(z)^{-1} |j\ket\bra j| M_1(z) 
\end{equation}

 belong to ${\mathfrak g }$, for $j=1, 2, \dots, 2M$
as they are the generators of the curves $D_j(t)$, $M_2(z)D_j(t)M_2^{-1}(z)$, $M_1^{-1}(z)D_j(t)M_1(z)$ which lie in $G$ as 
\begin{eqnarray*}
D_j(t)&=&D_j(t)M_2(z)M_1(z)(M_2(z)M_1(z))^{-1}, \\
 M_2(z)D_j(t)M_2^{-1}(z) &=&  M_2(z)D_j(t)M_1(z)(M_2(z)M_1(z))^{-1},\\
M_1^{-1}(z)D_j(t)M_1(z)  &=& (M_2(z)M_1(z))^{-1}M_2(z)D_j(t)M_1(z).
\end{eqnarray*}

The generators in (\ref{eq:generators}) have the same block structure as the $M_{j}$. We compute the relevant blocks. For $i M_2(z) |j\ket\bra j| M_2(z)^{-1}$ we get
\begin{eqnarray*}
\frac{i}{r^2}\begin{pmatrix}z & -t \cr t & -\bar{z} \end{pmatrix}
\begin{pmatrix}1&0\cr0&0\end{pmatrix}
\begin{pmatrix}\bar{z} & -t \cr t & -z \end{pmatrix}&=&
\frac{i}{r^2}\begin{pmatrix}1& -tz  \cr t\bar{z}  & -t^2\end{pmatrix}\\
\frac{i}{r^2}\begin{pmatrix}z & -t \cr t & -\bar{z} \end{pmatrix}
\begin{pmatrix}0&0\cr0&1\end{pmatrix}
\begin{pmatrix} \bar{z} & -t \cr t & -z\end{pmatrix}&=&
\frac{i}{r^2}\begin{pmatrix}-t^2& tz  \cr -t\bar{z}  & 1\end{pmatrix}.
\end{eqnarray*}
Similarly, for $i M_1(z)^{-1} |j\ket\bra j| M_1(z)$, the blocks take the form
\begin{eqnarray*}
\frac{i}{t^2}
\begin{pmatrix}z& r \cr r & \bar{z}\end{pmatrix}
\begin{pmatrix}1&0\cr0&0\end{pmatrix}
\begin{pmatrix}\bar{z}& -r \cr -r & z\end{pmatrix}&=&\frac{i}{t^2}
\begin{pmatrix}1& -rz \cr r\bar{z} & -r^2\end{pmatrix}\\
\frac{i}{t^2}
\begin{pmatrix}z& r \cr r & \bar{z}\end{pmatrix}
\begin{pmatrix}0&0\cr0&1\end{pmatrix}
\begin{pmatrix}\bar{z}& -r \cr -r & z\end{pmatrix}&=&\frac{i}{t^2}
\begin{pmatrix}-r^2& rz \cr -r\bar{z} & 1\end{pmatrix}.
\end{eqnarray*}

Now using these matrices for $z\notin\bR$ and the diagonal matrix $i |j\ket\bra j|$, $j=1,2$  one gets by taking suitable real linear combinations of the matrices above that, in both cases, the relevant blocks are  generated by 
\[\left\{ i 
\begin{pmatrix}1&0\cr0&0\end{pmatrix}, i 
\begin{pmatrix}0&0\cr0&1\end{pmatrix}, i 
\begin{pmatrix}0&1\cr-1&0\end{pmatrix}, 
\begin{pmatrix}0&1\cr1&0\end{pmatrix} \right\}.\]

For real $z$ use the curves
$D_j(t), D_eM_2D_j(t)M_2^{-1}D_e^{-1}, D_o^{-1}M_1^{-1}D_j(t)M_1D_o$, with 
$D_e=\oplus \left(
\begin{array}{cc}
  1&0      \\
  0& w       
\end{array}
\right)
$ and $D_o=\oplus \left(
\begin{array}{cc}
  w  &0      \\
  0& 1      
\end{array}
\right)$ for $w\in\bT$ which amounts to perform the change $z\mapsto w^{-1}z$.

  Taking into account the shift in the blocks and the period $2M$ of the indices in the matrices, we get that the restrictions of  ${\mathfrak g }$ and ${\mathfrak u}_{M}(1,1)$ to their tridiagonal elements, mod $2M$ coincide.

To go off the diagonals we use commutators, i.e. we exploit that $X, Y\in {\mathfrak g }$ implies $[X,Y]\in {\mathfrak g }$.

  Let $A_k=|k+1\ket\bra k|+|k\ket\bra k+1|\in  {\mathfrak g }$, for $k\in\bZ_{2M}$. Considering $[A_j,A_k]$ for all values of $j,k$, we generate a basis of all anti self-adjoint matrices that have non zero real matrix elements at distance two away from the diagonal (and in the corners, by periodicity). By commuting $A_k$ with $\tilde A_j=i(|j+1\ket\bra j|-|j\ket\bra j+1|)\in  {\mathfrak g }$, we get a basis of self-adjoint matrices with non zero purely imaginary elements on the  same upper and lower diagonals (plus corners) only. These matrices correspond to the restriction of all matrices in ${\mathfrak u}_M(1,1)$ to these diagonals.

  We generalize the argument as follows: Assume we already generated a basis of all matrices $A\in {\mathfrak u}_{M}(1,1)$ such that $A_{jk}=0$ if $|j-k|>m$, $m$ fixed. Again, periodicity is implicit here.

Let ${\mathfrak u}_M(1,1)\ni B_j^{\pm}(m)=|j+m\ket\bra j|\pm |j\ket\bra j+m|$.  We compute
\[[A_{j+m},B_j^{\pm}(m)]=B_j^{\mp}(m+1).\]

This way we generate all matrices $A\in {\mathfrak u}_M(1,1)$ such $A_{jk}=0$ if $|j-k|>m+1$.

  Hence  by induction, we see that ${\mathfrak g}={\mathfrak u}_M(1,1)$, so that $G=U_{M}(1,1)$.


\ep

\begin{lemma}\label{lem:irreducible}
$\tau\left(C^{\ast}GC\right)=\tau\left(Sp(M,\bC)\right)$ is $L_{2p}$ irreducible for $p\in\{1,\ldots,M\}$. 

{\bf Proof.} 
Denote $e_{i}, \quad i\in\{1,\ldots,4M\}$  the canonical basis vectors of $\bR^{4M}$.
By definition (see \cite{bl} with adaptation to our symplectic form)
\[L_{q}:=span\left\{v\in\Lambda^{q}\bR^{4M}; v=Me_{1}\wedge Me_{3}\wedge\ldots\wedge Me_{2q-1}, M\in Sp(2M,\bR)\right\}\]
\end{lemma}
for $q\le2M$. 
Remark that the  set of directions in $L_{q}$ corresponds to the set of isotropic subspaces of $\bR^{4M}$. 

$\tau\left(C^{\ast}G C\right)$ is $L_{q}$-irreducible if there is no proper linear subspace $V\subset L_{q}$ invariant under $\Lambda^{q}\tau\left(C^{\ast}G C\right)$.

Consider $M$ real numbers
\[a_{1}>a_{2}>\ldots a_{M}>1.\]
The $4M\times 4M$ diagonal matrix
\[A=diag\left(a_{1},\frac{1}{a_{1}},a_{2},\frac{1}{a_{2}},\ldots,a_{M},\frac{1}{a_{M}},a_{1},\frac{1}{a_{1}},a_{2},\frac{1}{a_{2}},\ldots,a_{M},\frac{1}{a_{M}}\right)\]
belongs to $\tau\left(C^{\ast}GC\right)$ and $e_{1}\wedge e_{3}\wedge\ldots\wedge e_{2q-1}$ is an eigenvector of $\left(\Lambda^{q}A\right)^{n}$ for all $n$ with simple dominant eigenvalue $>1$. Thus for an invariant subspace $V$ of $L_{q}$ either $e_{1}\wedge e_{3}\wedge\ldots\wedge e_{2q-1}\in V$ which implies $\Lambda^{q}M(e_{1}\wedge e_{3}\wedge\ldots\wedge e_{2q-1})\in V, \forall M$, thus $V=L_{q}$, or $e_{1}\wedge e_{3}\wedge\ldots\wedge e_{2q-1}\in V^{\perp}$ which implies for all $w\in V$
\[0=\left\bra\Lambda^{q}M^{t} w, e_{1}\wedge e_{3}\wedge\ldots\wedge e_{2q-1}\right\ket=
\left\bra w, \Lambda^{q}Me_{1}\wedge e_{3}\wedge\ldots\wedge e_{2q-1}\right\ket\]
thus $V^{\perp}=L_{q}\Longleftrightarrow V=\{0\}$. Thus we conclude the claimed irreducibility for $q=2p$.

\ep

\begin{lemma}\label{lem:contracting}
$\tau\left(C^{\ast}GC\right)=\tau\left(Sp(M,\bC)\right)$ is $2p$ contracting for $p\in\{1,\ldots,2M-1\}$.
\end{lemma}

{\bf Proof.} For any $a\in\bR\setminus 0$ there exist $x,y\in\bR$ with $x^{2}-y^{2}=1$ such that $\left(\begin{array}{cc}x& y  \\  y& x  \end{array}\right)$, which belongs to $U(1,1)$, has eigenvalues $a, 1/a$. Taking such matrices as blocks one sees that there exists an element of $U_{M}(1,1)$ whose singular values are distinct: $a_{1}>a_{2}>\ldots>a_{M}>1>1/a_{M}\ldots$ and thus an element of $\tau\left(C^{\ast}GC\right)$ with $2M$ distinct singular values
\[b_{1}=b_{2}=a_{1}>\ldots>b_{2p-1}=b_{2p}=a_{p}>\ldots>b_{2M-1}=b_{2M}=a_{M}>0.\]
Thus $b_{2p+1}^{n}/b_{2p}^{n}\to_{n\to\infty}0$ and it follows from proposition 2.1, p. 81 of \cite{bl} that $\tau\left(C^{\ast}GC\right)$ is $2p$ contracting.
\ep

\begin{remark} To summarize:  we have proved that if  the transfer matrices generate the complex symplectic group $Sp(M,\bC)$ then the results of \cite{bl} apply, i.e.: the Lyapunov spectrum is simple. The results in \cite{bl} are stated for real groups only. While it is remarked  in their introduction that these results should hold in the complex case, this seems not to be obvious to specialists in the field.
\end{remark}

\section{Thouless formula and  the mean Lyapunov exponent}\label{sec:thouless}
In this section we shall prove the announced identity in a series of lemmata. The quasienergy  will be called $z$ from now on.
\begin{theorem}\label{thm:thouless}
Let $M\in\bN$. For the first $M$ Lyapunov exponents associated with  $U$ defined in Definition (\ref{def:exponents}) it holds:
\[\frac{1}{M}\sum_{i=1}^{M}\lambda_{i}=\frac{1}{2}\log{\frac{1}{rt}}\ge\frac{1}{2}\log2
\]
\end{theorem}

{\bf Proof.}
Let $z\in\bT$. Denoting by $P_{2L}(z)$ the propagator
\begin{eqnarray*}
P_{2L}(z):\Omega&\to& U_{M}(1,1)\\
P_{2L}(z)(p)&:=&\Phi_{z}(L,p)\left(\Phi_{z}(-L,p)\right)^{-1}
\end{eqnarray*}

we have for $m\in\{1,\ldots,M\}$
\begin{equation}\label{eq:exterior}
\sum_{i}^{m}\lambda_{i}=\lim_{L\to\infty}\frac{1}{4L}\log{\Vert\wedge^{m}P_{2L}(z)(p)\Vert}\qquad p\  a.e. 
\end{equation}
where $\wedge^{m}$ denotes the m-th exterior product;  (c.f. \cite{a}, ch.3).

We analyse the above limit in proposition \ref{propo:thoulessFormula} below and show:
\[\frac{1}{M}\sum_{i}^{M}\lambda_{i}={2}\int_{\bT}\log\vert z-x\vert dl(x)+\frac{1}{2}\log\frac{1}{rt}.\]
The assertion follows by an explicit calculation proving that
\[\int_{\bT}\log{\vert z-x\vert} dl(x)=0.\]

\begin{propo}({Thouless formula})\label{propo:thoulessFormula}
Let $M\in\bN, z\in\bC\setminus 0$ then
\[\frac{1}{M}\sum_{i}^{M}\lambda_{i}(z)={2}\int_{\bT}\log\vert z-x\vert dl(x)+\frac{1}{2}\log\frac{1}{rt}-\log\vert z\vert\]
\end{propo}

 {\bf Proof:} Will be done in Appendix 1.
\hfill\ep

\begin{remark}Remark that we prove in particular that the density of states is the Lebesgue measure, see lemma \ref{thm:estimate} below.
\end{remark}

\subsection{Bounds on the localization length}\label{sec:localizationlength}

We now use the Thouless formula and an $M$ independent bound on the largest Lyapunov exponent to  derive a bound on  the localization length.  We remark that this bound is very crude and that more involved techniques should be established to get more detailed information; c.f. \cite{rs} and references therein. 

First observe that a lower bound on the mean  Lyapunov exponent together with a tight upper bound on the largest,  implies a lower bound on all.

\begin{lemma}\label{posly}
Let $\kappa>0, \delta>0$ such that $\forall M\in\bN, z\in\bT$

\[
\frac{1}{M}\sum_{j=1}^M\lambda_j\geq \kappa,\quad \mbox{and}\quad \lambda_1\leq \kappa+\delta, 
\]
then, for all $j=0,1,\ldots, M-1 $, 
\begin{equation}\label{eq:esb}
\lambda_{j+1}\geq \kappa-\frac{j\delta}{M-j}.
\end{equation}
\end{lemma}

{\bf Proof:}
First note that $\lambda_1\geq \frac1M\sum_{j=1}^M\lambda_j$. Thus $\lambda_1\geq \kappa$, 
which corresponds to (\ref{eq:esb}) for $j=0$. 
Similarly, using also the upper bound on $\lambda_1$, we have for any $1\leq j\leq M-1$,
\[
M\kappa\leq (\sum_{k=1}^{j}+\sum_{k=j+1}^{M})\lambda_k\leq j(\kappa+\delta)+\sum_{k=j+1}^{M}\lambda_k
\]
so that
\[
\lambda_{j+1}\geq \frac{1}{M-j}\sum_{k=j+1}^{M}\lambda_k\geq \kappa-\frac{j\delta}{M-j}.
\]
\hfill\ep

{\bf Remark:}

In view of localization properties, the estimate is useful only if 
\begin{equation}\label{slp}
\kappa>(M-1)\delta.
\end{equation}

\medskip

We now estimate the cocycle to derive an upper bound on the largest Lyapunov exponent, which is uniform in the 
quasienergy and width of the strip $M$.
\begin{propo}
Let $M\in\bN$ 
\begin{enumerate}
\item For the generator of the cocycle defined in (\ref{def:generator}) it holds
\[\Vert A(p)\Vert\le\frac{1}{rt}(1+r)(1+t);\]
\item it follows: $2\lambda_{1}\le\log\left(\frac{1}{rt}\right)+\log\left( (1+r)(1+t)\right)$.
\item There exists a $c>0$ such that for  $M\in\bN$ it holds:
\begin{eqnarray*}
dist(r, \{0,1\})<e^{-cM}&\Longrightarrow&\\
\xi_{M}=\frac{1}{\lambda_{M}}&\le&\frac{2}{\log\left(\frac{1}{rt}\right)-(M-1)\log\left((1+r)(1+t)\right)}.
\end{eqnarray*}

\end{enumerate}
\end{propo}
{\bf Proof.} 
The estimate on $A$ follows from its definition. The estimate on $\lambda_{1}$ is obtained using the equality (\ref{eq:exterior}). Finally, from the estimate (\ref{eq:esb}) it follows
\[\xi_{M}=\frac{1}{\lambda_{M}}\le\frac{2}{\log\left(\frac{1}{rt}\right)-(M-1)\log\left((1+r)(1+t)\right)}.\]
The bound is symmetric around $t=r=\frac{1}{\sqrt{2}}$ and finite for $r$ sufficiently away from the critical point $\frac{1}{\sqrt{2}}$ because of the singularity of $\log{1/rt}$.

\hfill\ep

\section{{Spectral Localization}}\label{sec:localization} 

We follow the strategy which was successfully employed for the case of one dimensional Schr\"odinger operators: polynomial boundedness of generalized eigenfunctions, positivity of the Lyapunov exponent and spectral averaging. We lean on the work of \cite{bhj, hjs}. Our result is: 

\begin{theorem}\label{thm:spectrallocalization}
Let $M\in{\mathbb N}$,  $rt\neq0$.
 Then, the Chalker Coddington model on the cylinder exhibits spectral localization throughout the spectrum, almost surely. More precisely, 
 
\begin{enumerate} 
\item the almost sure : spectrum $\Sigma$,  continuous spectrum $\Sigma_c$ and  pure point spectrum $\Sigma_{pp}$ of $U(p)$ satisfy
\[
\Sigma=\Sigma_{pp}={\bT} \qquad \mbox{and} \qquad \Sigma_{c}=\emptyset ;
\] 
\item the eigenfunctions decay exponentially, almost surely.
\end{enumerate}

\end{theorem}

{\bf Proof: } We prove the theorem in Appendix 2.

\section{Appendix 1}
We follow the strategy of \cite{cs} and first prove the lower bound

\begin{equation}\label{eq:lowerbound}
\frac{1}{M}\sum_{i}^{M}\lambda_{i}(z)\ge2\int_{\bT}\log\vert z-x\vert dl(x)+\frac{1}{2}\log\frac{1}{rt}-\log\vert z\vert
\end{equation}
for $0\neq z\in\bC\setminus\bT$
which follows from Lemma \ref{thm:estimate}  equation (\ref{eq:finiteLestimate}) below in the limit $L\to\infty$.
\hfill

\begin{lemma}
Denote $U^{D}$ the unitary defined by restriction of $U$ to \\ $l^{2}\left(\{-2L, \ldots, 2L\}, l^{2}(\bZ_{2M})\right)$ with reflecting boundary conditions:  the scattering picture for the links which are incoming to walls at $-(2L+1)$ and $2L+1$ reads
\[U^{D}e_{-2L, 2k+1}=e_{-2L,2k+2}, \quad U^{D}e_{2L, 2k}=e_{2L,2k+1}.\]

For $z\in\bC$ let
\[F_{z}:=\left\lbrace\psi\in l^{2}(\bZ_{2M}); \psi_{2k+1}=z \psi_{2k+2}, k\in\bZ_{M}\right\rbrace,\]
\[G_{z}:=\left\lbrace\psi\in l^{2}(\bZ_{2M}); z\psi_{2k+1}=\psi_{2k}, k\in\bZ_{M}\right\rbrace\]
and denote $Q_{F}$ the orthogonal projection to a subspace $F$. It holds: 
\begin{eqnarray*}
&{}&z \hbox{ is an eigenvalue of } U^{D}\\
&{}&\Longleftrightarrow
\ppsi_{2L}=P_{2L}(z)\ppsi_{-2L} \hbox{ and } \ppsi_{-2L}\in F_{z} \hbox{ and } \ppsi_{2L}\in G_{z}\\
&{}&\Longleftrightarrow
Ker\left(Q_{G_{z}^{\perp}}P_{2L}(z)Q_{F_{z}}\right)\neq\{0\}
\end{eqnarray*}
\end{lemma}

{\bf Proof.}
It holds
\[U^{D}\psi=z\psi\Longrightarrow\]
\[\psi_{-2L, 2k+1}=z\psi_{-2L, 2k+2} \hbox{ and } z\psi_{2L, 2k+1}=\psi_{2L, 2k }\]
so
\[\ppsi_{-2L}\in F_{z} \hbox{ and } \ppsi_{2L}\in G_{z}.\]
The identity
\[\ppsi_{2L}=P_{2L}(z)\ppsi_{-2L}\]
holds by construction of the transfer matrices so 
\[U^{D}\psi=z\psi\Longleftrightarrow 
Q_{G_{z}^{\perp}}
P_{2L}(z)Q_{F_{z}}\ppsi=0.\]
\hfill\ep

\begin{lemma}
Denote the ``even'' subspace of $ l^{2}(\bZ_{2M})$ by
\[E:=span\{e_{2k}; k\in\bZ_{M}\}.\]
For $z\neq0$ there exist invertible operators $V_{z}, W_{z}$ on $l^{2}(\bZ_{2M})$ such that $W_{z}(E)=F_{z}$ and $V_{z}(E)=G_{z}^{\perp}$ such that
\begin{enumerate}
\item 
\begin{eqnarray*}
z \hbox{ is an eigenvalue of } U^{D} \Longleftrightarrow\\
\det\left(Q_{E}V_{z}^{-1}P_{2L}(z)W_{z}Q_{E}\right)=0
\end{eqnarray*}
where we understand the determinant to apply to  the restriction to $E$.
\item For $z\neq0$; $\{z_{1},\ldots,z_{(4L+1)2M}\}$ the eigenvalues of $U^{D}$ it holds:
\begin{equation}
\label{eq:determinant}
\vert z\vert^{(4L+1)M}\vert \det\left(Q_{E}V_{z}^{-1}P_{2L}W_{z}Q_{E}\right)\vert=\frac{1}{(rt)^{2LM}}\Pi_{i=1}^{(4L+1)2M}\vert z-z_{i}\vert.
\end{equation}
\end{enumerate}
\end{lemma}

{\bf Proof.}
Fix $0\neq z\in\bC$. 

In the following  $N_{j}, D_{j}$ denote generic, $z$ independent matrices whose precise values may change from line to line. The $D_{j}$ are diagonal.

The transfer matrix $A_{z}$ defined in (\ref{def:generator}) is of the form
\[A_{z}=\frac{1}{rt}\left( z^{2}D_{1}Q_{O}+z^{-2}D_{2}Q_{E}+zN_{1}+N_{2}+\iz N_{3}\right)\]
where $O$ denotes the ``odd'' subspace defined by $O+E=l^{2}\left(\bZ_{2M}\right)$. Thus
\[(r t)^{2L} P_{2L}=z^{4L}D_{1}Q_{O}+z^{-4L}D_{2}Q_{E}+\sum_{j=-4L+1}^{4L-1}z^{j}N_{j}.\]
Note that
\[F_{z}=span\left\{\frac{1}{\sqrt{2}}\left(z e_{2k+1}+e_{2k+2}\right);k\in\bZ_{M}\right\}\]
\[G_{z}=span\left\{\frac{1}{\sqrt{2}}\left( e_{2k}+\iz  e_{2k+1}\right);k\in\bZ_{M}\right\}.\]
On $l^{2}(\bZ_{2M})$ define the operators 
\[W_{z}:=\frac{1}{\sqrt{2}}\sum_{k\in\bZ_{M}}| z e_{2k+1}+e_{2k+2}\rangle\langle e_{2k+2}|+|-e_{2k+1}+\iz e_{2k+2}\rangle\langle e_{2k+1}|\]
\[V_{z}:=\frac{1}{\sqrt{2}}\sum_{k\in\bZ_{M}}| - e_{2k+1}+z e_{2k}\rangle\langle e_{2k}|+|e_{2k}+\iz e_{2k+1}\rangle\langle e_{2k+1}|.\]
Then $W_{z}Q_{E}=Q_{F_{z}}W_{z}$ and $V_{z}Q_{E}=Q_{G_{z}^{\perp}}V_{z}$. Moreover one checks that
\begin{equation}
W_{z}^{2}=\bI,\qquad V_{z}^{-1}=KV_{z}K\label{eq:propertiesVW}
\end{equation}
with $K:=\sum_{k\in\bZ_{M}}|e_{2k+1}\rangle\langle e_{2k}|+|e_{2k}\rangle\langle e_{2k+1}|$.
It follows:
\[z \hbox{ is eigenvalue  of } U^{D}\Longleftrightarrow ker\left(Q_{E}V_{z}^{-1}P_{2L}W_{z}Q_{E}\right)\neq\lbrace0\rbrace.\]
Now 
\[Q_{E}V_{z}^{-1}P_{2L}W_{z}Q_{E}=\]
\[\frac{1}{2}\sum_{k,m}|e_{2k}\rangle\left\langle -e_{2k+1}+\frac{1}{\overline{z}}e_{2k}, P_{2L}\left(ze_{2m+1}+e_{2m+2}\right)\right\rangle\langle e_{2m+2}|=\]
\[\frac{1}{(rt)^{2L}}\left(z^{4L+1}D_{1}Q_{O}+z^{-4L-1}D_{2}Q_{E}+\sum _{\vert j\vert < 4L+1}z^{j}N_{j}\right).\]
Multiplication by $z^{4L+1}$ implies that for some $a_{j}\in\bC$
\[ z^{(4L+1)M}\det\left(Q_{E}V_{z}^{-1}P_{2L}W_{z}Q_{E}\right)=\frac{1}{(rt)^{2L M}}\sum_{0}^{(8L+2)M}z^{j}a_{j}.\]
$D_{1}$ is unitary thus,  in particular, $\vert a_{(8L+2)M}\vert=1$. $z^{(4L+1)M}\det\ldots $ being a polynomial of degree $(8L+2)M$ whose leading coefficient has modulus $(rt)^{-2L M}$ and which is zero on the $(4L+1)2M$ eigenvalues of $U^{D}$ the formula for the determinant follows.

\hfill\ep

We now  prove  convergence of  the finite volume ($L<\infty$) density of states   $\mu_L^M$  as $L\rightarrow\infty$  to a non random measure: the density of state. Then we show that this measure equals the Lebesgue measure.

\begin{lemma}\label{thm:estimate}
Denote $\mu_{L}^{(M)}$ the measure defined by
\[\frac{1}{(4L+1)2M} tr f(U^{D})=:\int_{\bT}f(x)d\mu_{L}^{M}(x)\qquad (f\in C(\bT)).\]

Then
\begin{enumerate}

\item \[\mu_{L}^{M}\to_{L\to\infty}^{vaguely}dl\]
the Lebesgue measure on $\bT$.
\end{enumerate}

\item For $M\in\bN$ there exists $c_{M}>0$ such that for all $L\in\bN, 0\neq z\in\bC\setminus\bT$

\begin{eqnarray}
&{ }&
\frac{1}{M}\frac{1}{4L}\log{\Vert \wedge^{M}P_{2L}\Vert}\nonumber\\
\label{eq:finiteLestimate}
&{ }&\ge\frac{1}{2}\log{\frac{1}{rt}}+(2+\frac{1}{2L})\int_{\bT}\log\vert z-x\vert d\mu_{L}^{M}(x)\\
&{ }& -\left(1+\frac{1}{4L}\right)\log\vert z\vert- \frac{c_{M}}{L}.\nonumber
\end{eqnarray}

\end{lemma}

{\bf Proof.} 
1.: We first prove the existence of a nonrandom limit measure. 
{
The first step consists in showing that $p$ a.e, 
\[\lim_{L\to\infty}\int_{\bT }f(U^D(p)) d\mu_{L}^{M}=\frac{1}{4}\left\{
\bE\left(\langle e_{0,0},f(U)e_{0,0}\rangle\right)
+
\bE\left(\langle e_{1,1},f(U)e_{1,1}\rangle\right)\right.\]
\[+\left.
\bE\left(\langle e_{1,0},f(U)e_{1,0}\rangle\right)+
\bE\left(\langle e_{0,1},f(U)e_{0,1}\rangle\right)\right\}
=:\int f d\mu_{\infty}^{M}, 
\] 
for all $ f\in C(\bT)$.
This follows from a classical argument based on ergodicity, separability of $C(\bT)$  and that $U-U^D$ has norm and rank uniformly bounded in $L$, see e.g. \cite{j} for the details for the unitary case. 
}

In order to identify $\mu_{\infty}^{M}$ recall that the normalized Lebesgue measure $dl$ on $\bT$ is uniquely characterized by :
\[\int_{\bT} p_{n} dl=\delta_{n,0}\qquad n\in\bZ\]
where $p_{n}(x):=x^{n}$. Consider the space of loops of euclidean length $n$ starting at $(0,0)$ :
\[\Gamma_{(0,0)}=\left\{\gamma:\{0,\ldots,n\}\to\{-2L,\ldots,2L\}\times\bZ_{2M},\gamma(0)=\gamma({n})=(0,0)\right\}. \]
Then because of the structure of $U$
\[
\langle e_{0,0},p_{n}(U)e_{0,0}\rangle=
\sum_{\gamma\in\Gamma_{(0,0)}}
\langle e_{0,0},Ue_{\gamma(1)}\rangle\ldots
\langle e_{\gamma(n-1)},Ue_{0,0}\rangle.
\]
Now $\langle e_{\gamma(j)},U(p)e_{\gamma(j+1)}\rangle=l(p)t^{\alpha}r^{\beta}$ for some $\alpha, \beta\ge0$ and $l$ a uniformly distributed random variable. Thus
$\bE\left(\langle e_{0,0},p_{n}(U)e_{0,0}\rangle\right)=\delta_{n,0}$. Applying the same argument to $\langle e_{1,1},f(U)e_{1,1}\rangle, \langle e_{0,1}\ldots$ we conclude:
\[d\mu_{\infty}^{M}=dl\]

2.:  By formula (\ref{eq:determinant}):
\[\frac{1}{4LM} \log\left\vert\det\left(Q_{E}V_{z}^{-1}P_{2L}W_{z}Q_{E}\right)\right\vert =\]
\[\frac{1}{2}\log\frac{1}{rt}+ (2+\frac{1}{2L})\int_{\bT}\log\vert z-x\vert d\mu_{L}^{M}(x)-\left(1+\frac{1}{4L}\right)\log\vert z\vert\le\]
\[\frac{1}{4LM}\log\Vert\wedge^{M}\left(Q_{E}V_{z}^{-1}P_{2L}W_{z}Q_{E}\right)\Vert\le\]
\[\frac{1}{M}\frac{1}{4L}\log\Vert\wedge^{M}P_{2L}\Vert+\frac{1}{L}\underbrace{\frac{1}{4M}\left(\log\Vert\wedge^{M}Q_{E}V_{z}^{-1}\Vert+\log\Vert\wedge^{M}W_{z}Q_{E}\Vert\right)}_{=:c_{M}}\]
where we used the identity
\[\det Q_{E}AQ_{E}=
\left\langle e_{0}\wedge\ldots\wedge e_{2M-2},\wedge^{M}A e_{0}\wedge\ldots\wedge e_{2M-2}\right\rangle.\]
From this the claim follows:

\hfill\ep

We turn now to the proof of the opposite inequality: 

\begin{equation}\label{eq:upperbound}
\frac{1}{M}\sum_{i}^{M}\lambda_{i}(z)\le2\int_{\bT}\log\vert z-x\vert dl(x)+\frac{1}{2}\log\frac{1}{rt}-\log\vert z\vert
\end{equation}
for $0\neq z\in\bC\setminus\bT:$

\begin{propo}\label{cle} Suppose that for any choice of sets of vectors $\left\{d_0^-,d_2^-,\ldots ,d_{2M-2}^-\right\}$ and $\left\{d_0^+,d_2^+,\ldots ,d_{2M-2}^+\right\}$ in the "odd" subspace $O$
\begin{align}
&\limsup_{L\rightarrow \infty}\frac{1}{M(4L+1)}\times\nonumber\\
&\times\log |\left\bra (e_0+d_0^-)\wedge \ldots \wedge (e_{2M-2}+d_{2M-2}^-),\right. \nonumber\\
&\left.\hspace{3cm}\wedge^M (V_z^{-1}P_{2L}(z)W_z) (e_0+d_0^+)\wedge \ldots \wedge (e_{2M-2}+d_{2M-2}^+)\right\ket|\nonumber\\
 &\leq \frac{1}{2}\ln \frac{1}{rt} + 2\int_{\mathbb T}\log |x -z|\, dl(x)-\log |z|\label{eq:hypothesis}\enspace
\end{align}
then, for all $0\neq z \in {\mathbb C}\setminus\bT,$
$$\frac{1}{M}\sum_{i}^{M}\lambda_{i}(z)\le \frac{1}{2}\log \frac{1}{rt}+ 2\int_{\mathbb T}\log|x-z|\, dl(x) -\log |z|$$
\end{propo}
{\bf Proof: } The vectors of the form $\{(e_0+d_0)\wedge \ldots \wedge (e_{2M-2}+d_{2M-2}); d_0,\ldots, d_{2M-2} \in O\}$ span $\wedge^M {\mathbb C}^{2M}$. On the other hand, given any spanning sets $S_1$ and $S_2$ in $\wedge^M {\mathbb C}^{2M}$, the mapping $\|\cdot \|_S$ defined by
\[
\|A\|_S \equiv\sup_{\phi\in S_1,\psi\in S_2}|\bra\phi, A\psi \ket|
\]
defines a norm over the algebra of operators in $\wedge^M {\mathbb C}^{2M}$. It follows that there exists $c>0$, which depends on $S_1$ and $S_2$, such that for any matrix $A$, $\|A\|_S\geq c\|A\|$, hence that
$$\frac{1}{M}\sum_{i}^{M}\lambda_{i}(z)\leq \frac{1}{2} \log \frac{1}{rt} +2\int_{\mathbb T}\log |\zeta-z|\, dl(\zeta)-\log |z|\, .$$
\ep

We now prove that the inequality (\ref{eq:hypothesis}) is satisfied. This will be achieved in two steps. In order to keep track of the $L$ dependence denote by $U^{D}_{L}$ the former $U^{D}$. Now reinterpret  the left hand side of (\ref{eq:hypothesis})  as the characteristic polynomial of a deformation of $U^{D}_{L}$ denoted $V^{D}_{L}$; more precisely: we aim at equation (\ref{eq:deformation}) below. The problem is then reduced to the proof of the weak convergence of the associated sequence of counting measures $(\nu_{L,z}^M)$ towards $\mu^M$. 

\subsection{Deformation of $U^{D}_{L}$}

Let $L$ in ${\mathbb N}$ and define the matrix $V^{D}_{L+1}$ on $l^2(\{-2L-2, \ldots, 2L+2\}, \l^2({\mathbb Z}_{2M}))$ by: $\forall \psi\in l^2(\{-2L-2, \ldots, 2L+2\}, l^2({\mathbb Z}_{2M})),$
\begin{eqnarray*}
(V^{D}_{L+1}\psi)_{2L+2,2k} &=& \sum_{l=0}^{M-1}B^+_{2k,2l}\psi_{2L,2l}\\
(V^{D}_{L+1}\psi)_{2L+1,k} &=& \psi_{2L+1,k}\\
(V^{D}_{L+1}\psi)_{2L,2k+1} &=& \sum_{l=0}^{M-1}C^+_{2k+1,2l+1}\psi_{2L+2,2l+1}\enspace,
\end{eqnarray*}
\begin{eqnarray*}
(V^{D}_{L+1}\psi)_{-2L,2k} &=& \sum_{l=0}^{M-1}B^-_{2k,2l}\psi_{-2L-2,2l}\\
(V^{D}_{L+1}\psi)_{-2L-1,k} &=& \psi_{-2L-1,k}\\
(V^{D}_{L+1}\psi)_{-2L-2,2k+1} &=& \sum_{l=0}^{M-1}C^-_{2k+1,2l+1}\psi_{-2L,2l+1}\enspace,
\end{eqnarray*}
with the same reflecting boundary conditions as $U^{D}_{L+1}$ and\\ $(V^{D}_{L+1}\psi)_{\mu,\nu}=(U^{D}_{L+1}\psi)_{\mu,\nu}$ for any values of $(\mu,\nu)$ which were not described previously. The matrix $V^{D}_{L+1}$ is a deformation of the matrix $U^{D}_{L+1}$, but its structure remains close to the structure of $U^{D}_{L+1}$. Note that $span\{e_{2L+1,k},e_{-2L-1,j}; j,k\in\bZ_{2M}\}$ belongs to $ker\left( V_{L+1}^{D}-\bI\right)$. For $\psi$  an eigenvector of $V^{D}_{L+1}$ associated with the eigenvalue $z$, 
$$V^{D}_{L+1}\psi= z\psi.$$
This implies that either $\psi\in span\{e_{2L+1,k},e_{-2L-1,j}; j,k\in\bZ_{2M}\}$ and $z=1$, or $\psi_{-2L-2}\in F_{z}$ and $\psi_{2L+2}\in G_{z}$ and
\begin{eqnarray*}
\psi_{2L} &=& P_{2L}(z)\psi_{-2L}\\
\psi_{2L+2} &=& A^+(z)\psi_{2L}= z^{-1}Q_E B^+ Q_E \psi_{2L} + z Q_O {C^+}^{-1} Q_O \psi_{2L}\\
\psi_{-2L} &=& A^-(z)\psi_{-2L-2}= z^{-1}Q_E B^- Q_E \psi_{-2L-2} + z Q_O {C^-}^{-1} Q_O \psi_{-2L-2}\enspace.
\end{eqnarray*}
The transfer matrices $A^+(z)$ and $A^-(z)$ are deformations of the matrices $A_z$. This construction is useful to establish the following lemma. In the following, $z$ will be fixed as a parameter.

\begin{lemma} Let $(d_{2k}^+)_{k\in \{0,\ldots, M-1\}}$ and $(d_{2k}^-)_{k\in \{0,\ldots, M-1\}}$ two families of vectors belonging to the "odd" subspace $O$. These families are the columns of two corresponding matrices denoted $D^+$ and $D^-$ respectively. Assume that for $z\neq0$, $\max(\|D^-\|,\|D^+\|)\leq \max(\frac{1}{\vert z\vert},\vert z\vert)$ and consider the matrix $V^{D}_{L}$ parametrized by $z$ 
\begin{eqnarray*}
B^+_{z} &=& z+TD^+\\
C^+_{z} &=& z(1- z^{-1}D^+ T)^{-1}\\
B^-_{z} &=& z+z^2 D^{- \ast}\\
C^-_{z} &=& (z^{-1}-D^{- \ast})^{-1}
\end{eqnarray*}
with
\[
T = \left(
\begin{array}{cccccc}
0 & & && 1\\
1 & 0 && &\\
& 1&\ddots& &\\
& & && \\
& & &1 & 0
\end{array}
\right)\enspace.
\]
Then,
\begin{align}
&\bra (e_0+d_0^-)\wedge \ldots \wedge (e_{2M-2}+d_{2M-2}^-), \nonumber\\
&\hspace{2cm}\wedge^M (V_z^{-1}P_{2L}(z)W_z) (e_0+d_0^+)\wedge \ldots \wedge (e_{2M-2}+d_{2M-2}^+)\ket\nonumber\\
&=\bra e_0\wedge \ldots \wedge e_{2M-2}, \wedge^M (V_z^{-1}A^-P_{2L}(z)A^+ W_z) e_0\wedge \ldots \wedge e_{2M-2}\ket \enspace.\label{eq:deformation}
\end{align}
\end{lemma}
{\bf Proof:} By (\ref{eq:propertiesVW}): $W_z^2=I$, $V_z^{-1}=K V_z K$ , $z\neq 0$. It follows for all $k\in \{0, \ldots, M-1\}:$
\begin{eqnarray*}
A^+ W_z e_{2k} &=& W_z (e_{2k}+d_{2k}^+)\\
A^{- \ast} V_z^{-1 \ast}e_{2k} &=& V_z^{-1 \ast} (e_{2k}+d_{2k}^-)\enspace.
\end{eqnarray*} 
\hfill\ep

Given $z$, $D^+$ and $D^-$ and the associated matrix $V^{D}_{L+1}$, we consider the corresponding eigenvalue problem:
$$V^{D}_{L+1}\psi = z' \psi$$
The complex number $z'$ is an eigenvalue of $V^{D}_{L+1}$ iff
\begin{gather*}
(z'-1)^{4M}\bra e_0\wedge \ldots \wedge e_{2M-2}, \wedge^M (V_{z'}^{-1}A^-_{z}(z')P_{2L}(z')A^+_{z}(z') W_{z'}) e_0\wedge \ldots \wedge e_{2M-2}\ket =0\enspace
\end{gather*}
where $A^{\pm}_{z}(z')=  z'^{-1}Q_E B^\pm_{z} Q_E  + z' Q_O {C^\pm_{z}}^{-1} Q_O$.

Once multiplied by ${z'}^{(4L+5)M}$, the left-hand side is a polynomial of degree $2M(4L+5)$ in $z'$. Following the Thouless argument, we get for the  logarithm of the modulus divided by $4ML$:
\begin{gather*}
\frac{1}{2}\log\frac{1}{rt} -\log |z'| + (2+\frac{1}{2L})\int_{{\cal B}(0,R_z)}\log |x-z'|\, d\nu^M_{L,z}(x)+O(\frac{1}{L})\enspace,
\end{gather*}
where the family of measures $\nu^M_{L,z}$ are supported on some closed ball ${\cal B}(0,R_z)$, due to the fact that $\sup_L \|U^{D}_{L}-V^{D}_{L}\| < \infty$. Note that if $z'=z$, $A^+_{z}(z)=A^+$ and $A^-_{z}(z)=A^-$.

\subsection{End of proof of inequality (\ref{eq:hypothesis})}

We split the proof in the two following lemmas, whose proof is an adaptation of the argument given in \cite{cs}. 

\begin{lemma} If $(\nu_L)_{L\in {\mathbb N}}$ and $\mu$ are measures supported on ${\cal B}(0,R)$ for some $R>0$, and if $(\nu_L)$ converge weakly to $\mu$, then for any $z\in {\mathbb C}$
\[
\int_{\mathbb T}\log|\zeta-z|\, d\nu_L(\zeta)\leq \int_{\mathbb T}\log|\zeta-z|\, d\mu(\zeta)\enspace.
\]
\end{lemma}
\noindent {\bf Proof: } Given $z\in {\mathbb C}$, let $f_{\epsilon}$ be defined by:
\begin{gather*}
\left\{\begin{array}{l}
f_{\epsilon}(\zeta)= \log |\zeta-z|\, \mbox{if}\, |\zeta-z|\geq \epsilon\\
f_{\epsilon}(\zeta)= \log |\epsilon|\, \mbox{if}\, |\zeta-z|\leq \epsilon
\end{array}\right.
\end{gather*}
Since the support is compact,
\[
\lim_{L \rightarrow \infty}\int_{\mathbb T}f_{\epsilon}(\zeta)\, d\nu_L(\zeta)= \int_{\mathbb T}f_{\epsilon}(\zeta)\, d\mu(\zeta)\enspace.
\]
On the other hand, for any $\zeta$ in ${\mathbb T}\, ,$
$$\log |\zeta-z|\leq f_{\epsilon}(\zeta)$$
so that:
\begin{eqnarray*}
\limsup_{L\rightarrow \infty}\int_{\mathbb T}\log |\zeta-z|\, d\nu_L(\zeta) &\leq & \limsup_{L\rightarrow \infty}\int_{\mathbb T}f_{\epsilon}(\zeta)\, d\nu_L(\zeta)\\
&=& \int_{\mathbb T}f_{\epsilon}(\zeta)\, d\mu(\zeta).
\end{eqnarray*}
The result follows by monotone convergence theorem when $\epsilon$ goes to zero.\ep

\noindent{\bf Remark:} Let us note that the $\ast$-algebra of trigonometric polynomials ${\cal F}$ defined by:
\[
{\cal F}= \{f\in C({\cal B}(0,R));f(r,\theta)=\sum_{k_1+|k_2|\leq N} a_{k_1,k_2} r^{k_1}e^{ik_2 \theta}, N\in {\mathbb N}_0, k_1\in {\mathbb N}_0, k_2\in {\mathbb Z}\}
\]
separates  points and contains the constants. Its closure under the supremum norm is $C({\cal B}(0,R))$. The weak convergence of the measures is equivalent to have for all $f$ in ${\cal F},$
\[
\lim_{L\rightarrow \infty}\int_{{\cal B}(0,R)} f(\zeta)\, d\nu_L(\zeta) = \int_{{\cal B}(0,R)} f(\zeta)\, d\mu(\zeta)\enspace.
\]

\begin{lemma} As a Borel measure on ${\mathbb C}$, the sequence of measures $(\nu^M_{L,z})$ para\-metrized by $z,M$ converges almost surely weakly to $dl$ as $L$ tends to infinity.
\end{lemma}

{\bf Proof: } Let $(r_{j,z} e^{\xi_{j,z}})_{j=1}^{2M(4L+1)}$ and $(e^{\lambda_j})_{j=1}^{2M(4L+1)}$ be the eigenvalues of the problems with reflecting boundary conditions for $V$ and $U_D$ which correspond respectively to the modified and unmodified "potentials". We have that:
\begin{eqnarray*}
\nu^M_{L,z} &=& \frac{1}{2M(4L+1)} \sum_j \delta_{r_{j,z} e^{\xi_{j,z}}}\\
\mu^M_L &=& \frac{1}{2M(4L+1)} \sum_j \delta_{e^{\lambda_j}}\enspace.
\end{eqnarray*}
These measures are supported on some ${\cal B}(0,{R_z})$. We will drop the $z$ subscript in the sequel. Since we already know that $(\mu^M_L)$ converges almost surely weakly to $dl$, we only need to show that for any nonnegative integer $k_1$ and any integer $k_2,$
\[
\lim_{L\rightarrow \infty}\frac{1}{2M(4L+1)}\sum_{j=1}^{2M(4L+1)} e^{ik_2 \lambda_j}- r_j^{k_1}e^{ik_2 \xi_j}=0\enspace.
\]
Actually, it is enough to prove it for nonnegative integers $k_1$, $k_2$. Let us fix such a couple $(k_1,k_2)$ and decompose the term on the left-hand side as follows:
\begin{eqnarray*}
\frac{1}{2M(4L+1)}\sum_{j=1}^{2M(4L+1)} e^{ik_2 \lambda_j}- r_j^{k_1}e^{ik_2 \xi_j} &=& T_1(L)+T_2(L)\\
\mbox{where}\quad T_1(L) &=& \frac{1}{2M(4L+1)}\sum_{j=1}^{2M(4L+1)} (r_j^{k_2}- r_j^{k_1})e^{ik_2 \xi_j}\\
T_2(L) &=& \frac{1}{2M(4L+1)} \mbox{Tr} (U_D^{k_2}-V^{k_2})\enspace.
\end{eqnarray*}
If $k=\min(k_1,k_2)$ and $l=max(k_1,k_2)$, we have
\[
|T_1(L)|\leq \frac{1}{2M(4L+1)}\sum_{j=1}^{2M(4L+1)}r^k|r_j^{l-k}-1| \leq \frac{R_z^k}{2M(4L+1)}\sum_{j=1}^{2M(4L+1)}|r_j^{l-k}-1|\enspace.
\]
Following \cite{cs}, we first prove that:
\[
\lim_{L\rightarrow \infty} \frac{1}{2M(4L+1)}\sum_{j=1}^{2M(4L+1)}|r_j-1|=0\enspace.
\]
We know that there exists two orthonormal bases $(\phi_j)_{j=1}^{2M(4L+1)}$ and $(\phi_j')_{j=1}^{2M(4L+1)}$ such that:
\[
V^{D}_{L}=\sum_{j=1}^{2M(4L+1)}\mu_j(V^{D}_{L})|\phi_j'\ket\bra \phi_j|\enspace,
\]
where $(\mu_j(V^{D}_{L}))$ are the singular values of the operator $V^{D}_{L}$. Actually, $\mu_j(V^{D}_{L})=r_j$ and we assume them to be ordered: $\mu_{j+1}(V^{D}_{L})\geq \mu_j(V^{D}_{L}) \geq 0$. Note that:
$$\{\mu_j^2(V^{D}_{L});j\in \{1, \ldots, 2M(4L+1)\}\}=\sigma({V^{D}_{L}}^{\ast}V^{D}_{L})\setminus \{0\}.$$
Since for each $j\in \{1, \ldots, 2M(4L+1)\}$, $\mu_j(U^{D}_{L})=1$ we deduce from the remark following Theorem 1.20 in \cite{s} that:
\[
\sum_{j=1}^{2M(4L+1)}|r_j-1| = \sum_{j=1}^{2M(4L+1)}|\mu_j(V^{D}_{L})-\mu_j(U^{D}_{L})| \leq \sum_{j=1}^{2M(4L+1)}\mu_j(V^{D}_{L}-U^{D}_{L})\enspace.
\]
Since $V^{D}_{L}-U^{D}_{L}$ has rank and norm uniformly bounded in $L$, we obtain that:
\[
\lim_{L\rightarrow \infty}T_1(L)=0\enspace.
\]
The term $T_2(L)$ will be treated  in a similar way. The operator $U^{D}_{L}-V^{D}_{L}$ has rank and norm uniformly bounded in $L$. This implies that for all integer $k_2$, ${U^{D}_{L}}^{k_2}-{V^{D}_{L}}^{k_2}$ has also rank and norm uniformly bounded in $L$. So,
\[
\lim_{L\rightarrow \infty}T_2(L)=0\enspace,
\]
which concludes the proof. \ep

The above lemmata together with equation (\ref{eq:deformation}) establish the inequality (\ref{eq:hypothesis}) which implies  (\ref{eq:upperbound}). We finish with the proof of the Thouless formula on $\bT$:

\begin{lemma}\label{cle2} For all $z\in {\mathbb T},$ $$\frac{1}{M}\sum_{i}^{M}\lambda_{i}(z)= \frac{1}{2} \log\frac{1}{rt} + 2 \int_{\mathbb T}\log|x-z|\, dl(x)$$
\end{lemma}
{\bf Proof: } We note with \cite{cs0} that 
$\lim_{L\rightarrow \infty}\frac{1}{4L}\log \| \wedge ^M (P_{2L})(z)\|$ is subharmonic in $\bC\setminus \{0\}$  and $\int_{\bT} \log |z-x| dl(x)$  subharmonic on $\bC\setminus \bT$. The two  exceptional sets are of measure zero in $\bC$, these quantities must agree everywhere.  
\ep
\begin{remark}
We note that the above proof does not depend on the specific form of the density of states.
\end{remark}
\section{Appendix 2} 
Now we prove theorem \ref{thm:spectrallocalization} in several steps.

By theorem \ref{thm:finite} the localization length is finite for all values of the parameters.
Note that the spectrum is characterized by the existence of generalized eigenfunctions: 

Suppose that the support of $E_{p}(\cdot)$, the spectral resolution of
$U(p)$, is the whole circle $\bT$.

\begin{propo}\label{simonrevu} For $M\in\bN$, $p\in\Omega$ the spectrum of $U(p)$ is the closure of the set
\[ S_{p} = \{ z\in \bT ; U(p)\phi = z\phi \mbox { has a non-trivial polynomially bounded solution} \}
\]
 and $E_p(\bT \setminus S_{p}) =0$.
\end{propo}

{\bf Proof: } 
The stated behaviour at infinity of the generalized eigenvectors and the spectrum of $U(p)$ are related by
Sh'nol's Theorem. This well known deterministic fact for self-adjoint operators was proven in \cite{bhj}  to hold  in the unitary setup for band matrices on $l^2(\bZ)$. It is straightforward to check that the result holds for band matrices on $l^2(\bZ, \bC^{2M})$, with $M$  finite.\ep

Secondly we prove the existence of a finite cyclic subspace: 

\begin{lemma}\label{cyc}
Let $M\in\bN, rt\neq0$.
Denote ${I_0} := {\{0\}}\times\bZ_{2M}$. The vectors $\{e_{\mu}; \mu\in {I_0}\}$ span a cyclic subspace of $l^{2}\left(\bZ\times\bZ_{2M}\right)$.
\end{lemma}

{\bf Proof:} The only non vanishing elements in $U$ are the blocks given in equation (\ref{def:block}). Denoting generically the elements of $S$ by
\[
S=:\left(
\begin{array}{cc}
  \alpha &  \beta   \\
  \gamma  & \delta  
\end{array}
\right)
\]
and observing that $U^{-1}_{\mu,\nu}=\overline{U_{\nu,\mu}}$ we have
\[
\left(
\begin{array}{cc}
  U_{(2j+1,2k);(2j,2k)} &  U_{(2j+1,2k);(2j+1,2k+1)}   \\
 U_{(2j,2k+1);(2j,2k)}  & U_{(2j,2k+1);(2j+1,2k+1)}   
\end{array}
\right)=\left(
\begin{array}{cc}
  \alpha &  \beta   \\
  \gamma  & \delta  
\end{array}
\right)\]
\[=\left(
\begin{array}{cc}
  U_{(2j+2,2k+2);(2j+2,2k+1)} &  U_{(2j+2,2k+2);(2j+1,2k+2)}   \\
 U_{(2j+1,2k+1);(2j+2,2k+1)}  & U_{(2j+1,2k+1);(2j+1,2k+2)}   
\end{array}
\right)\]
and
\[
\left(
\begin{array}{cc}
  U^{-1}_{(2j,2k);(2j+1,2k)} &  U^{-1}_{(2j,2k);(2j,2k+1)}   \\
 U^{-1}_{(2j+1,2k+1);(2j+1,2k)}  & U^{-1}_{(2j+1,2k+1);(2j,2k+1)}   
\end{array}
\right)=\left(
\begin{array}{cc}
  \overline{\alpha} &  \overline{\gamma}   \\
  \overline{\beta}  & \overline{\delta}  
\end{array}
\right)\]
\[=\left(
\begin{array}{cc}
  U^{-1}_{2j+2,2k+1;2j+2,2k+2} &  U^{-1}_{2j+2,2k+1;2j+1,2k+1}   \\
 U^{-1}_{2j+1,2k+2;2j+2,2k+2}  & U^{-1}_{2j+1,2k+2;2j+1,2k+1}   
\end{array}
\right).\]
Computing $Ue_{(0,2k)}=\alpha e_{(1,2k)}+\gamma e_{(0,2k+1)}$ and the corresponding expressions for $U^{-1}e_{(1,2k)}, Ue_{(0,2k+1)}, U^{-1}e_{(-1,2k+1)}$ we infer: 
\begin{eqnarray*}
e_{(1,2k)}=\frac{1}{\alpha}\left(Ue_{(0,2k)}-\gamma e_{(0,2k+1)}\right)\\
e_{(1,2k+1)}=\frac{\beta}{\alpha}e_{(0,2k)}-\frac{\gamma}{\overline{\beta}}U^{-1}e_{(0,2k+1)}\\
e_{(-1,2k+1)}=\frac{1}{\gamma}\left( Ue_{(0,2k+1)}-\alpha e_{(0,2k+2)}\right)\\
e_{(-1,2k+2)}=\frac{\delta}{\gamma}e_{(0,2k+1)}-\frac{\alpha}{\overline{\delta}}U^{-1}e_{(0,2k+2)}.
\end{eqnarray*}
Thus vectors with indices in $\{\pm1\}\times\bZ_{2M}$ belong to the subspace generated by ${U^{\pm 1}(I_0)}$. The lemma follows by induction.
\ep

\medskip

Let $I= \{0,1\}\times\bZ_{2M} $, $\overline{\Omega}= \bT^{ {\mathbb Z}^{4M} \setminus I}$,
$\overline{\bP} = \otimes_{k\in  {\mathbb Z}^{4M}\setminus I}\ dl$,
$\overline{p} = \{\overline {p}_j\}_{j\in {\mathbb Z}^{4M}\setminus I}\in 
\overline{\Omega}$, and $\Theta_I=\{\theta_j\}_{j\in I}$. 
We shall use the notation 
$\Omega\ni p=(\overline{p}, \Theta_I)$.

Denote
\[
\lambda_M(p, z):=\lim_{L\rightarrow \infty }(\frac{1}{4L}\log(\|
\wedge^M P_{2L}(z)(p)\|) -\frac{1}{4L}\log(\|\wedge^{M-1} P_{2L}(z)(p)\|) ),
\]
if the limit exists. 

By construction, \ref{def:exponents}, it holds for almost every  $p$
\[
\lambda_M=\lambda_M(p)
\]

By definition $\lambda_M(p,z)$ is independent
of the finitely many $\Theta_I$, if $p=(\overline{p}, \Theta_I)$. By theorem \ref{thm:finite} there exists $\overline{\Omega}(z)\subset
\overline{\Omega}$ with
$\overline{\bP}(\overline{\Omega}(z))=1$ such that  for any
$z\in\bT\setminus\bR$
\begin{equation*}
\label{poslap} \lambda_M((\overline{p}, \Theta_I),z) 
=\lambda_M>0, 
\end{equation*} 
for all $\theta_{j}\in \Theta_I$ and all $\overline{p} \in
\overline{\Omega}(z)$. We can apply Fubini  to the measure
$\overline{\bP} \times dl $ to get the existence of
$\overline{\Omega}_0 \in \overline{\Omega}$ with
$\overline{\bP}(\overline{\Omega}_0)=1$ such that for every
$\overline{p} \in \overline{\Omega}_0$ there is
$B_{\overline{p}} \in \bT$ with $l(B_{\overline{p}}) =0$
and 
\begin{equation} \label{poslap2} \lambda_M((\overline{p}, \Theta_I),z) 
> 0 \ \ \ \mbox{for all $\theta_{j}\in \Theta_I,$ and all 
$z \in {B_{\overline{p}}}^C$}. 
\end{equation}

Then we show that for $\overline{p}\in\overline{\Omega_{0}}$, ${B_{\overline{p}}}^C$ is a support of the
spectral resolution of $U({(\overline{p}, \Theta_I)})$ for 
almost every $\theta_j\in \Theta_I$ w.r.t. $d^{|I|}l$ on $\bT^{|I|}$.

For any fixed $j\in I$, we introduce the spectral measures 
$\mu_{p}^j$ associated
with $U(p)=\int_{\bT}x\ d E_p(x)$ defined
for  all Borel sets $\Delta\in\bT$ by 
\begin{equation*}
\mu_{p}^{j}(\Delta)=\bra e_j | E_{p}(\Delta) | e_j\ket. 
\end{equation*}
Since $U(p)=D(p) \dS$, where $D(p)$ is diagonal, the variation 
of a random phase at one site is
described by a rank one perturbation. More precisely, dropping the
variable $p$ temporarily, we define $\widetilde{D}$ by taking
$\theta_j=1$ in the definition of $D$: 
\begin{equation*}\label{dhat}
\widetilde{D}=D +|e_j\ket\bra
e_j|(1-\theta_j)=e^{\log (\overline{\theta}_j)|e_j\ket\bra e_j|} D, 
\end{equation*} 
so that, with the obvious notations,
\begin{equation*}\label{uhat} 
\widetilde{U}=\widetilde{D}\dS=e^{\log (\overline{\theta}_j)|e_j\ket\bra e_j|} U. 
\end{equation*}
The unitary version of the spectral averaging formula, 
see \cite{c} and \cite{b}, reads in our case: 
for any $f\in L^1(\bT)$, 
\begin{equation*}
\int_\bT \ dl(\theta_j) \int_\bT f(x)
d\mu_{(\overline{p},\Theta_{I})}^{j}(x)
=\int_\bT f(x)\ dl(x). 
\end{equation*} 
Applied to $f=\chi_{B_{\overline{p}}}$, the
characteristic function of $B_{\overline{p}}$, this yields 
\begin{equation} 0 =
l(B_{\overline{p}}) = \int_\bT 
\mu_{(\overline{p},\Theta_{I})}^j(B_{\overline{p}})\, {dl(\theta_j)}.
\end{equation}
Consequently,
\begin{equation*}\label{star}
\mu_{(\overline{p},\Theta_{I})}^j(B_{\overline{p}}) =0, \ \ \
\mbox{for every $\theta_{k}\in \Theta_I, \ k\neq j$
and Lebesgue-a.e.\ $\theta_j$}.
\end{equation*} 
Therefore, for all $\overline{p}\in \overline{\Omega}_0$,
there exists $J_{\overline{p}}\subset \bT^{|I|}$ s.t.
$l({J_{\overline{p}}}^C)=0$ and 
\begin{equation} \label{final}
\Theta_I\subset J_{\overline{p}} \Rightarrow
\mu_{(\overline{p},\Theta_I)}^{j}(B_{\overline{p}})=0, \ \ \forall j\in I. 
\end{equation}

Now fix $\overline{p} \in \overline{\Omega}_0$ and $\Theta_I \subset J_{\overline{p}}$ and consider $p =
(\overline{p}, \Theta_I)$. By Lemma \ref{cyc} and (\ref{final}) 
we deduce that
$E_{p}(B_{\overline{p}}) = 0$. If $S_{p}$ is the
set from Sh'nol's Theorem~\ref{simonrevu},  then
the set $S_{p} \cap {B_{\overline{p}}}^C$ 
 is a support for
$E_{p}(\cdot)$.

Now take $z \in S_{p} \cap {B_{\overline{p}}}^C$. By
Theorem~\ref{simonrevu}, $U(p)\psi=z\psi$ has a
non-trivial polynomially bounded solution $\psi$. On the other
hand, by (\ref{poslap2}), $\lambda_M(p, z)>0$. Thus,
by Osceledec's Theorem, every solution which is polynomially
bounded necessarily has to decay exponentially both 
at $+\infty$ and $-\infty$, and therefore
it is  an eigenfunction of $U({p})$. In other words,
every $z \in S_{p} \cap {B_{\overline{p}}}^C$ 
is an eigenvalue of
$U(p)$, hence $S_{p} \cap {B_{\overline{p}}}^C$ is countable.
Therefore $E_{p}(\cdot)$ has countable support thus $U(p)$ has pure point spectrum. With
$$
\Omega_0 := \{ (\overline{p}, 
\{\theta_{j}\}_{j\in I})\ \ \mbox{s.t.} \ \ \overline{p} \in \overline{\Omega}_0, \ \{\theta_{j}\}_{j\in I}\subset \Theta_I \, \subset J_{\overline{p}} \}, 
$$
 we have
\begin{equation}
\label{nosing} p \in \Omega_0  \ \ \Rightarrow \ \ \sigma_{c}(U({p})) = \emptyset  . 
\end{equation}

Also, from $l({J_{\overline{p}}}^C) =0$ we have 
\begin{equation} \label{posmeasure} 
(\otimes_{j\in I} \, dl)(J_{\overline{p}}) = (\otimes_{j\in I} \, dl)
(\bT^{|I|})=1. 
\end{equation}

As $\overline{\bP}(\overline{\Omega}_0)=1$, we conclude from
(\ref{nosing}) and (\ref{posmeasure}) that 
\begin{equation*}
\bP(\sigma_{c}(U({p})) = \emptyset) \ge \bP(\Omega_0) =
\int_{\overline{\Omega}_0} d\overline{\bP}(\overline{p}) (
\otimes_{j\in I} \, dl) (J_{\overline{p}}) =1, 
\end{equation*}
which proves that 
$U({p})$ has almost surely pure point spectrum. The fact that the support of the density of state coincides with the almost sure spectrum, see \cite{j}, shows that $\Sigma_{pp}=\bT$.

We finally show that almost surely all eigenfunctions decay
exponentially. Note that we actually have shown above
that the event ``all eigenvectors of $U({p})$ decay at the
rate of the smallest Lyapunov exponent'' has probability one, since this
is true for all $p\in \Omega_0$. Measurability of this event was proven 
for the case of ergodic one-dimensional Schr\"odinger operators by 
Kotani and Simon in Theorem~A.1 of \cite{KoSi}. The proof of this
fact provided in \cite{KoSi} carries over to the CC model as well. It is enough 
to note that, due to Lemma~\ref{cyc}, we may use $\rho_{p}
= \sum_{j\in I}\mu_{p}^{j}$ as spectral measures in
their argument.
\ep

\section*{Acknowledgments}We should like to thank the referee for his constructive criticism and H. Schulz Baldes and H. Boumaza for enlightening discussions.
We  acknowledge gratefully support from the
grants Fondecyt Grant 1080675; Anillo PBCT-ACT13; MATH-AmSud, 09MATH05; Scientific Nucleus
Milenio ICM P07-027-F.


\begin{thebibliography}{xxxxxxx}
 
\bibitem[A]{a}Arnold, L.: Random dynamical systems. Springer Monographs in Mathematics. Springer-Verlag, Berlin, (1998)

\bibitem[ASS]{ass}Avron, J. E., Seiler, R., Simon, B.: Charge deficiency, charge transport and comparison of dimensions.  {\it Comm. Math. Phys.}  {\bf 159} , 399--422,  (1994).

\bibitem[B]{b} Bourget, O.: Singular continuous Floquet operator for periodic
Quantum systems,{\it J. Math. Anal. Appl.} {\bf 301}, 65-83,
(2005).

\bibitem[BHJ]{bhj} Bourget, O., Howland, J.S., Joye, A.: Spectral Analysis of Unitary
Band Matrices, {\it Commun. Math. Phys.} {\bf 234}, 191-227
(2003).

\bibitem[BL]{bl} Bougerol, P. and Lacroix, J.: Products of random matrices with applications to
              {S}chr\"odinger operators,  Progress in Probability and Statistics, {\bf 8}, Birkh\"auser Boston (1985)


\bibitem[Bou]{bou} Boumaza, H.: H\"older continuity of the IDS for matrix-valued Anderson models, {\it Rev. Math. Phys.} {\bf 20},  873-900 (2008).

\bibitem[BS]{bs} Boumaza, H., Stolz, G.: Positivity of Lyapunov exponents for Anderson-type models on two coupled strings, {\it Elec. J. Diff. Eq.} {\bf 47}, 11-18
(2007).

\bibitem[BESB]{besb}Bellissard, J., van Elst, A., and Schulz-Baldes, H.: ``The noncommutative geometry of the quantum Hall effect,'' J. Math. Phys. \textbf{35}, 5373-5451 (1994).
\bibitem[C]{c} Combescure, M.: Spectral Properties of a Periodically Kicked
Quantum Hamiltonian, {\it J. Stat. Phys.} {\bf 59}, 679-690,
(1990).

\bibitem[CC]{cc} Chalker, J.T., Coddington, P.D.: Percolation, quantum tunneling and the integer Hall effect, {\it J. Phys. C} {\bf 21}, 2665-2679, (1988).

\bibitem[CdVP]{cdvp1} Colin de Verdi\`ere, Y., Parisse, B.: \'Equilibre instable en r\'egime semi-classique. I. Concentration microlocale, {\it  Comm. Partial Diff. Equat.} {\bf 19},1535--1563, (1994).

\bibitem[CFKS]{cfks} Cycon, H.L., Froese, R.G., Kirsch, W., Simon,
B.: {\it Schr\"odinger Operators}, Springer Verlag, 1987.
\bibitem[CKM]{CKM} Carmona, R., Klein A., Martinelli, F.: Anderson
localization for Bernoulli and other singular potentials, {\it
Commun. Math. Phys.} {\bf 108}, 41-66, (1987).
\bibitem[CL]{cl} Carmona, R., Lacroix, J.:  {\it Spectral theory of random
    Schrodinger Operators}, Birkh\"auser, 1990.
    
\bibitem[CS1]{cs0} Craig, W. and Simon, B: Subharmonicity of the Lyaponov index, {\it Duke Math. J.} {\bf 50},  551-560, (1983)
    
\bibitem[CS2]{cs} Craig, W.; Simon, B.: Log H\"older continuity of the integrated density of states for stochastic Jacobi matrices.
Comm. Math. Phys. 90 (1983), no.2, 207-218. 

\bibitem[FH]{fh} Fertig, H.A., Halperin, B.I.: Transmission coefficient of an electron through a saddle-point potential in a magnetic field, {\it Phys. Rev. B} {\bf 36}, 7969--1976, (1987).

\bibitem[G]{g} Graf, G. M.: Aspects of the integer quantum Hall effect. Spectral theory and mathematical physics: a Festschrift in honor of Barry Simon's 60th birthday, 429--442, Proc. Sympos. Pure Math., {\bf 76}, Part 1, Amer. Math. Soc., Providence, RI, (2007)

\bibitem[GKS]{gks} Germinet, F., Klein, A., Schenker, J.: Dynamical delocalization in random Landau Hamiltonians, {\it Ann. of Math.} {\bf 166}, 215--244, (2007).

\bibitem[GM]{gm} Gol{}dshe{\u\i}d, I. Ya. and Margulis, G. A.: Lyapunov exponents of a product of random matrices, {\it Uspekhi Mat. Nauk} {\bf 44}, 13--60, (1989).



\bibitem[HS]{hs} Helffer, B., Sj\"ostrand, J.: Analyse semi-classique pour l'\'equation de Harper , {\it M\'emoires de la S.M.F.} {\bf 34}, 1--113, (1988).

\bibitem[HJS]{hjs} Hamza, E., Joye, A., Stolz, G.: { Localization for Random Unitary Operators}, 
{\it Lett.  Math. Phys.}, {\bf 75},  255-272, (2006).
\bibitem[J1]{j} Joye, A.: Density of States and Thouless Formula
for Random Unitary Band Matrices, {\it Ann. Henri Poincar\'e} {\bf 5},
347--379, (2004).

\bibitem[KOK]{kok} Kramer, B., Ohtsuki, T., Kettemann, S.: Random network models and quantum phase transitions in two dimensions, {\it  Phys. Rep.} {\bf 417}, 211--342, (2005).  


\bibitem[KS]{KoSi} Kotani, S., Simon, B.: Localization in general
one-dimensional random systems, {\it Commun. Math. Phys.} {\bf
112}, 103--119, (1987).

\bibitem[RS]{rs} Roemer, R., Schulz-Baldes, H.: 
Random phase property and the Lyapunov spectrum for disordered multi-channel systems preprint, (2009). http://de.arxiv.org/abs/0910.5808

\bibitem[S]{s} Simon B.: Trace ideals and their applications, Second edition. Mathematical Surveys and Monographs, 120. American Mathematical Society, (2005)


\bibitem[T]{t} Trugman, S.A.: Localization, percolation, and the quantum Hall effect, {\it Phys.
Rev. B} {\bf 27}, 7539--7546, (1983).

\bibitem[${\rm TKN^{2}}$]{tknn}Thouless, D.J., Kohmoto, M., Nightingale, M.P., den Nijs, N.: Quantized Hall Conductance in a Two-Dimensional Periodic Potential, {\it Phys. Rev. Lett.} {\bf 49}, 405--408, (1982).

\bibitem[W]{w2} Wang, W.M.: Microlocalization, Percolation, and Anderson Localization for the Magnetic Schr\"odinger Operator with a Random Potential, {\it Journal of Funct. Anal.} {\bf 146}, 1--26, (1997).
\end{thebibliography}
\end{document}